\documentclass[manuscript]{aastex}
\usepackage{natbib}
\slugcomment{to appear in Astronomical Journal}

\shorttitle{Chemical composition of the open cluster M6 (NGC\,6405)}
\shortauthors{K{\i}l{\i}\c{c}o\u{g}lu et al.}

\newcommand{\vmic}{\ensuremath{v_{\mathrm{mic}}}}
\newcommand{\vrad}{\ensuremath{v_{\mathrm{rad}}}}
\newcommand{\vsini}{\ensuremath{v_{{\mathrm{e}}}\sin i}}
\newcommand{\teff}{\ensuremath{T_\mathrm{eff}}}
\newcommand{\logg}{\ensuremath{\mathrm{log}\ g}}
\newcommand{\loggf}{\ensuremath{\mathrm{log}\ gf}}
\newcommand{\kms}{\ensuremath{\mathrm{km\,s^{-1}}}}

\newcommand{\hbeta}{\textit{H}$_\beta$}
\newcommand{\Msol}{\ensuremath{\,\mathrm{M_{\odot}}}}
\newcommand{\Kelvin}{\ensuremath{\,\mathrm{K}}}

\begin{document}


\title{Chemical composition of intermediate mass stars members of the M6 (NGC\,6405) open cluster}


\author{T. K{\i}l{\i}\c{c}o\u{g}lu\altaffilmark{}}
\affil{Ankara University, Faculty of Science, Department of Astronomy and Space Sciences, 06100, Tando\u{g}an, Ankara, Turkey}
\email{tkilicoglu@ankara.edu.tr}

\author{R. Monier\altaffilmark{1}}
\affil{LESIA, UMR 8109, Observatoire de Paris Meudon, Place J. Janssen, Meudon, France}
\email{Richard.Monier@obspm.fr}

\author{J. Richer\altaffilmark{}}
\affil{D\'epartement de physique, Universit\'e de Montr\'eal, 2900, Boulevard Edouard-Montpetit, Montr\'eal QC, Canada  H3C 3J7}
\email{Jacques.Richer@umontreal.ca}

\author{L. Fossati\altaffilmark{}}
\affil{Argelander-Institut f\"ur Astronomie der Universit\"at Bonn, Auf dem H\"ugel 71, 53121, Bonn, Germany}
\email{lfossati@astro.uni-bonn.de}

\and

\author{B. Albayrak\altaffilmark{}}
\affil{Ankara University, Faculty of Science, Department of Astronomy and Space Sciences, 06100, Tando\u{g}an, Ankara, Turkey}
\email{balbayrak@ankara.edu.tr}


\altaffiltext{1}{Laboratoire Lagrange UMR 7293, Universit\'e de Nice Sophia, Nice, France}


\begin{abstract}


We present here the first abundance analysis of 44 late B, A and F-type members of the young open cluster M6 (NGC 6405, age about 75 Myrs). Low and medium resolution spectra, covering the 4500 to 5800\,\AA{} wavelength range, were obtained using the FLAMES/GIRAFFE spectrograph attached to the ESO Very Large Telescopes (VLT). We determined the atmospheric parameters using calibrations of the Geneva photometry and by adjusting the $H_{\beta}$ profiles to synthetic ones. 
The abundances of up to 20 chemical elements, from helium to mercury,  were derived for 19 late B, 16 A and 9 F stars by iteratively adjusting synthetic spectra to the observations. We also derived a mean cluster metallicity of $\mathrm{[Fe/H]=0.07\pm0.03}$ dex
from the iron abundances of the F-type stars
. We find that, for most chemical elements, the normal late B and A-type stars exhibit larger star-to-star abundance variations than the F-type stars do
probably because  of the faster rotation of the B and A stars. The abundances of
C, O, Mg, Si and Sc  appear to be anticorrelated to that of Fe, while the opposite holds for the abundances of Ca, Ti, Cr, Mn, Ni, Y, and Ba about as expected if radiative diffusion is efficient in the envelopes of these stars. In the course of this analysis, we discovered five new peculiar stars: 
one mild-Am, one Am, and one Fm star (HD 318091, CD-32\,13109, GSC\,07380-01211), one HgMn star (HD\,318126), and one He-weak P-rich (HD\,318101) star.
We also discovered a new spectroscopic binary, most likely a SB2.
We performed a detailed modelling of HD\,318101,the new He-weak P-rich CP star, using the Montr\'eal stellar evolution code XEVOL which treats self-consistently all particle transport processes. Although the overall abundance pattern of this star is properly reproduced, we find that detailed abundances (in particular the high P excess) resisted modelling attempts even when a range of turbulence profiles and mass loss rates were considered. Solutions are proposed, which are still under investigation.
 
\end{abstract}


\keywords{stars: abundances --- open clusters: individual(M6) --stars individual: HD 318101}



\section{Introduction}

Abundance determinations of late-B, A and F dwarfs in open clusters aim at elucidating the mechanisms of mixing at play in the interiors of these main-sequence stars.
Indeed open clusters are excellent laboratories to test stellar evolution as the stars members of an open cluster originate from the same original insterstellar material and have the same initial chemical composition and age.

This paper follows two series of paper addressing the chemical composition of F, A and late B-type dwarfs in open clusters and/or moving groups of different ages: i) \citet{varmon99}, \citet{mon05}, \citet{gebmon08}, \citet{gebetal08}, and \citet{gebetal10} for the  \object{Hyades}, the \object{Ursa Major Moving Group}, \object{Coma Berenices}, and the \object{Pleiades}, ii) \citet{fosetal07}, \citet{fosetal08}, and \citet{fosetal11} for \object{Praesepe} and \object{NGC 5460}. A review can be found in \citet{fos14}.

The  main objective of these works 
is twofold: first, we wish to improve our knowledge of the chemical composition of late-B, A and F dwarfs, and second, we aim at using these determinations to set constraints on particle transport processes in self consistent evolutionary models. Other groups also performed abundance analysis of intermediate-mass stars in open clusters, with a variety of goals \citep{stuetal06,viletal09,pacetal10,panetal10,jacetal11,carpan11,caretal14}.

In this paper, we present the first abundance analysis of 44 late B, A and F-type dwarfs members of the young open cluster M6.
\object{M6} (\object[NGC 6405]{NGC\,6405}) is an interarm object lying between the local arm and the Sagittarius arm \citep{vle74}. It is a moderately-rich cluster \citep{vanhag75} and it contains about 120 most-likely members (for $V\le15.1$) according to \citet{ant72}. Previous age, distance, color excess and metallicity determinations for the cluster are collected in Table \ref{intro}, along with our determinations (see Sec.\,\ref{metallicity} and \ref{hrd}). Early estimates of the distance, reddening and age of the cluster were obtained photometrically by \cite{rohetal59}\footnote{In this study, we use the identification number of the stars in the region of M6 from \cite{rohetal59}.} who derived a distance of 630 pc from photoelectric UBV observations of 132 stars, a mean reddening E(B-V) of 0.156\,mag and estimated an age of 100\,Myr. In a similar manner, \citet{egg60} derived a distance of 505\,pc from photometric observations of 66 stars and a mean reddening E(B-V)=0.13 mag. After a careful removal of field stars, \citet{vle74} has derived a colour excess E(B-V)=0.15\,mag, a distance of 450\,pc and an age of about 100\,Myr. \citet{sch85} performed Str\"omgren photometry of M6 up to V=12.0 mag, and derived a redening of E(b-y) = 0.110. \citet{maisch84} reported the detection of 3 chemically peculiar stars, with star number \object[V970 Sco]{77} exhibiting one of the strongest 5200 \AA{} depression ever measured. \citet{cam85} used the extinction corrected ultraviolet excess of stars in a color-color diagram to estimate the average metallicity, obtaining $\mathrm{[Fe/H]\sim0.07}$ dex. 

In this work, we have derived the abundances of up to 20 chemical elements, from Helium to Mercury,  for 19 B, 16 A and 9 F-type stars observed with the GIRAFFE spectrograph in the wavelength range from 4500 up to 5840\, \AA. 
The spectral types of these stars range from B5 (.ie a mass of about 4.3 $M_{\odot}$ ) to F6
 (1.4 $M_{\odot}$). We have looked for star-to-star variations of the elemental abundances, in particular for the B and A-type stars, and searched for putative correlations of the various abundances with effective temperature, projected rotational velocity and the iron abundance. 
In the framework of radiative diffusion, the abundances of a few elements, in particular manganese, are expected to increase with effective temperature
and should remain insensitive to rotation up to 120 $\kms$ since the timescale of diffusion are much shorter than those of rotational mixing \citep{charbonneaumi91}.
In the course of this abundance analysis, we have discovered five new chemically peculiar stars and a double-lined spectroscopic binary.

The target selection, membership assessment, and data reduction are described in section \ref{observations}. The determination of the effective temperatures and surface gravities (\teff{} and \logg{}), and abundance analysis are discussed in section \ref{analysis}. We discuss our results in Sections \ref{results} and \ref{discussion}. In section 5, we provide a detailed modeling of HD 318101, one of the newly found CP star, using the Montr\'eal stellar evolution code XEVOL.
Conclusions are gathered in Section \ref{conclusions}. 

\section{Observations, Data Reduction and Target Selection} \label{observations}

The spectra of 104 objects in the region of M6 were obtained using the Fibre Large Array Multi Element Spectrograph (FLAMES) instrument with the GIRAFFE spectrograph attached to the Unit 2 Kueyen of the Very Large Telescopes (VLT) at the European Southern Observatory (ESO) by one of us (LF). The instrument settings and target selection were carried out in the same way as described by \citet{fosetal11}.

Three different gratings were chosen for GIRAFFE: LR3, HR09B, and HR11. The low resolution (R=$\lambda$/$\Delta\lambda$=7500) LR3 setting, covering the 4490--5080\,\AA{} wavelength range, includes the \hbeta{} line and several strong lines of neutrals and singly ionized elements. The wide wavelength range of these single-order spectra encompasses the entire \hbeta{} line with at least 200\,\AA{} extension on each side of the \hbeta{} line, allowing a proper continuum normalisation of the Balmer line. The two medium resolution (R$\sim$25000) HR09B and HR11 settings, covering respectively the 5140--5380 and 5450--5750\,\AA{} wavelength ranges, were also adopted to derive abundances of further elements and with higher accuracy. In addition, the higher resolution spectra allow for a more precise determination of the projected rotational velocity (\vsini{}).

The spectra were reduced using the standard GIRAFFE data processing pipeline, which includes scattered light correction\footnote{$\mathrm{http://www.eso.org/observing/dfo/quality/GIRAFFE/pipeline/pipe\_reduc.html}$}. 
In order to correct for the sky background, we averaged the 4--6 sky spectra which have been obtained closest to each star, and substracted the average sky spectrum from each corresponding stellar spectrum. Each star was observed twice, six times, and three times with the LR3, HR09B, and HR11 settings, respectively, on May 30th and June 5th, 2007. We co-added the spectra obtained with the same gratings in order to increase the signal-to-noise ratio (S/N),after correcting  for the barycentric velocity correction. Using the spectra obtained with the HR09B setting, we looked for radial velocity variations (which were measured using cross-correlation with appropriate synthetic spectra). In the event we found significant radial velocity variations, we shifted the spectra to the same reference frame before co-addition. The observing log is given in Table \ref{log}.

The spectra were normalized to selected continuum points, ie. regions we knew free of lines from our synthetic spectra calculations,  adjusting a smooth spline function through these points. In the event that no true continuum could be found (e.g., for rapidly rotating stars or for the cooler stars), pseudo-continuum windows were selected and the location of the true continuum retrieved using synthetic spectra convolved for the appropriate rotation rate. All the spectra acquired with the HR11 setting show broad absorption features, much wider than the neighbouring stellar absorption lines. These broad features are located at 5705.65\,\AA{}, 5780.54\,\AA{}, and 5797.16\,\AA{} and correspond to three well-known diffuse interstellar bands (DIBs).

Since most members of the M6 open cluster still lack a radial velocity determination, their membership has yet to be firmly established. In order to select the most-likely cluster members, we first estimated effective temperatures, projected rotational velocities, and radial velocities by performing a preliminary spectral analysis, in a similar way as described in Sect.~\ref{analysis}. We primarily considered 56 targets with $\teff{} > 6100$\,K and spectra with S/N${}\sim100$, and selected 44 targets to perform a more thorough spectral analysis considering the membership from three criteria: i) proper motions, ii) radial velocity, and iii) position of the stars in the Hertzsprung-Russell (HR) diagram (Table \ref{fundamental}).

According to \citet{frimaj08}, only six stars (nos.\,1 [\object{BM Sco}], 17, 20, \object[HD 160221]{72}, 77 [\object{V970 Sco}], and HD\,318111) have been identified as cluster members from proper motions and radial velocity, with an averaged radial velocity of $-8.27\pm0.45\,\kms{}$. As a first criterion, we considered the stars for which we derived a radial velocity which differs by more than $\pm 5\,\kms{}$ from this average value as non-members. We then checked the membership probability of the remaining stars against \citet{diaetal06} and \citet{frimaj08}, who derived membership probabilities for stars in the field of view of M6 using high precision proper motions from the UCAC catalogue \citep{zacetal04} and the Tycho-2 catalogue, respectively. As a second criterion, we adopted as members the remaining stars having a membership probability larger than 25\% according to either \citet{diaetal06} or \citet{frimaj08}. We finally checked the locations of the remaining stars in the HR diagram (see Sec. \ref{hrd} for details). The stars which were found to lie on the average age isochrone were considered as members. Applying these three criteria yields a final list of 44  certain members (ie. with a membership probability of 100 \%).

We have searched various sources for possible binarity of the 44 retained members.
According to the photometric studies of M6 by \citet{ant72}, \citet{maisch84}, and \citet{sch85},  none of the 44 retained stars is known to be a binary . Inspection of the CCDM catalogue \citep{domnys95} reveals that HD\,160167 (no.\,115) is a double star. The difference in radial velocity between our measurements and those of \citet{frimaj08} supports the binary nature of HD\,160167. However, we could not find any evidence of the lines of the secondary in our spectra. This star may be a SB1 and we analysed it as  a single star.

\subsection{The Hertzsprung-Russell diagram of the cluster} \label{hrd}

\begin{figure}
\epsscale{.80}
\plotone{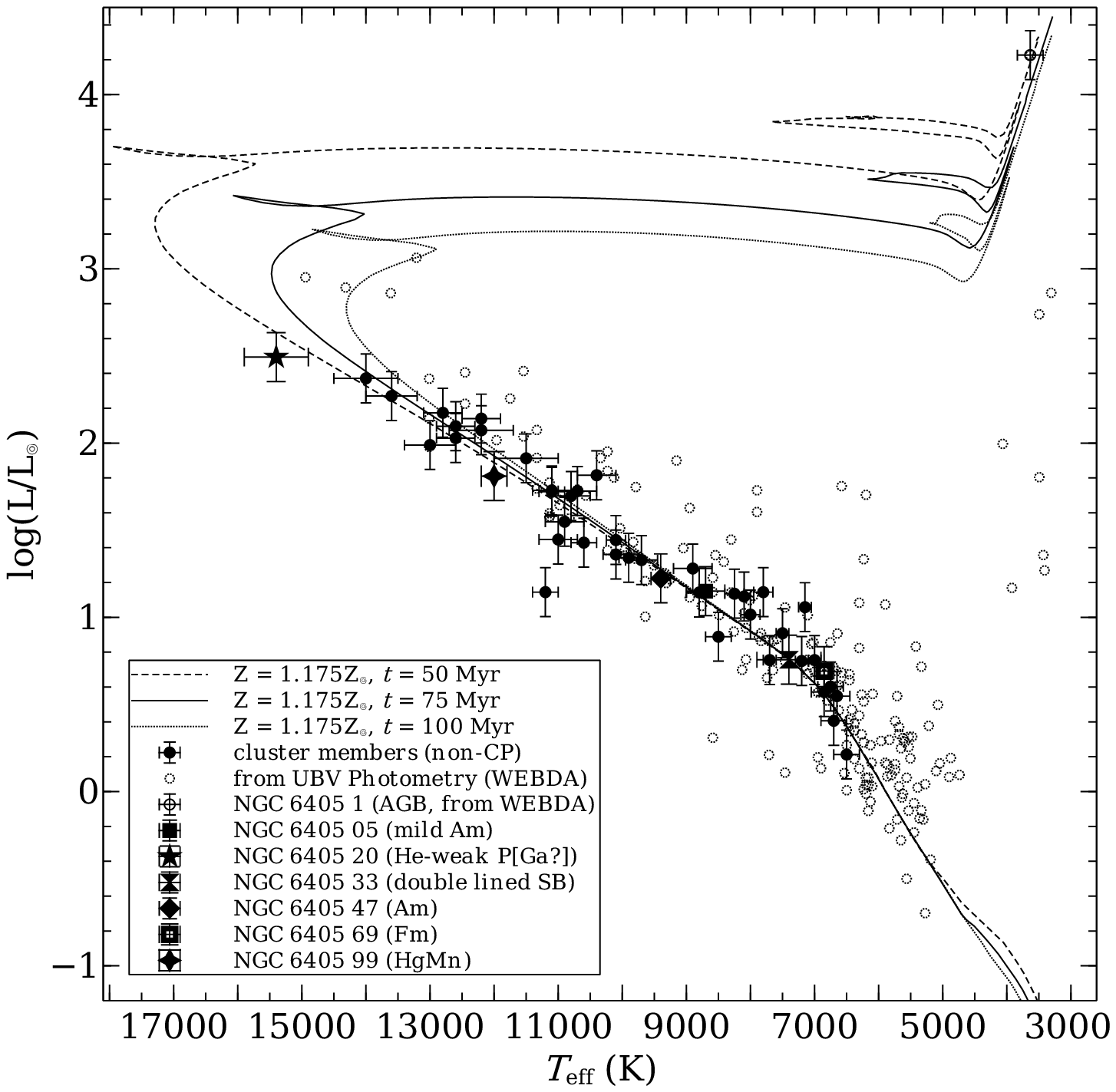}
\caption{HR diagram of the observed stars.}
\label{HRD}
\end{figure}

The Hertzsprung-Russell (HR) diagram of M6 is shown on Fig.\,\ref{HRD} displaying luminosities versus against the effective temperatures determined in section 3.2.1. We retrieved the Johnson UBV photometry of the stars from the WEBDA database. The luminosities were derived from the V magnitudes and the bolometric corrections from \citet{besetal98}, and an atmospheric extinction ($A_V=0.47\pm0.06$) calculated from :

$$A_V = 3.14E(B-V)\quad\hbox{\citep{schwie75}}.$$

We adopted a color excess $E(B-V)$ of $0.15\pm0.01$ (see Table \ref{intro}).
The uncertainties of the distance of the cluster ($\pm50$ pc), the first coefficient in the formula of \citet[$\pm0.10$]{schwie75}, and the color excess ($\pm0.01$) yield a typical uncertainty on $\log {L \over L_{\odot}}$ of about  $\pm0.14$. 
The luminosities are derived from the total integrated flux (estimated from the V magnitudes corrected with the bolometric corrections) and the distance $d$ by: $L =
4 \pi d^{2} F$.

In the HR diagram of Fig.\,\ref{HRD}, the  normal B, A and F-type stars are depicted as filled circle. Special symbols depict the five chemically peculiar stars and one double-lined SB. which we have discovered in the course of this analysis. We have not corrected the luminosities of binaries since we did not know the contribution of the secondary to the total flux.

We estimated the distance and age of the M6 cluster by adjusting  isochrones retrieved from the PARSEC\footnote{http://stev.oapd.inaf.it/cgi-bin/cmd} library \citep{breetal12}. This site allows to compute isochrones specifically  for the found cluster metallicity, which is $Z_{\mathrm{cluster}}=1.175Z_{\odot}$ (see sec. \ref{metallicity}) for M6. We thus derived a distance of $400\pm50$\,pc and an age of $75\pm25$\,Myr for M6. We have collected the luminosities, masses, and fractional ages of the members of M6 (following Landstreet et al., 2007) in Table\,\ref{fractional}.

\section{Abundance analysis} \label{analysis}

\subsection{Model atmospheres, spectrum synthesis and atomic data}

Model atmospheres were computed for the fundamental parameters of each star assuming solar composition \citep{gresau98} using the Linux version of the ATLAS9 code \citep{kur93,sboetal04,sbo05}. These models assume a plane-parallel geometry, Hydrostatic Equilibrium (HE), Radiative Equilibrium (RE), and Local Thermodynamical Equilibrium (LTE). We used the new opacity distribution function (ODF) tables described in \citet{caskur03}. For stars cooler than 8500\,K, convection was included adopting a mixing length parameter ($\alpha$) of 0.5 for 7000\,K $\le$ \teff{} $\le$ 8500\,K, and 1.25 for \teff{} $<$ 7000\,K according to the prescriptions in \citet{sma04}. Each model has 72 layers and was converged up to $\log\tau=-6.875$ usually within 15 iterations. For the five stars we identified as new Chemically Peculiar stars, we have computed ATLAS12 model atmospheres for the first set of peculiar abundances found with ATLAS9. New abundances were derived by adjusting synthetic spectra to the observed ones and the process was iterated until the abundances derived from spectral synthesis and those included in ATLAS12 were the same. Convergence was in generally met in two iterations. The termperature profile $T(\tau_{5000})$ in these ATLAS12 models differs little from the ATLAS9 models as the elements whose abundances were adjusted are not very abundant elements.

Abundances were derived by adjusting grids of synthetic spectra convolved with the appropriate rotation and instrumental profile to the normalized observed spectra since several of the retained stars rotate fairly fast and we deal with two different resolutions. The synthetic spectra were calculated using SYNSPEC48 \citep{hublan95} and its SYNPLOT interface. We adopted the solar abundances from \citet{gresau98}. 
We used Procyon (F5V) \citep{grigri79} , whose fundamental parameters are accurately known and whose composition is about solar \citep{ste85}, as a control star for the spectrum synthesis.
While attempting to synthesize the  observed Sc\,{\sc ii} lines of Procyon,
we realized that SYNSPEC48 produces systematically too strong Sc\,{\sc ii} lines. \citet{freetal05} already reported on this problem for a previous version of SYNSPEC. As \citet{freetal05} did, we incorporated the PFSAHA routine of ATLAS9 relative to Scandium into SYNSPEC48 and used this modified code (hereafter refered to as SYNSPEC48mod).  The synthetic spectra were broadened by a gaussian profile corresponding to the  appropriate instrument resolving power and by a parabolic profile for the appropriate \vsini{} of the star using the routine ROTIN3 provided along with SYNSPEC48 (this follows equation 18.14 in Gray (2005))
We modified the IDL graphic interface SYNPLOT and implemented a routine which searches for the minimum  of the  $\chi^2$  between the synthetic spectrum and the observed spectrum within a wavelength range specified by the user. 

We created an atomic linelist by using first the file {\sc gfall.dat} from Kurucz's database\footnote{http://kurucz.harvard.edu/linelists.html}, restricted to the 4490--5850\,\AA{} wavelength interval. This list includes line data mostly from the literature for the light and the heavy elements with critically evaluated transition probabilities of \citet{maretal88} and \citet{fuhetal88} and computed data by \citet{kur92} for the iron group elements. In the event more recent and more accurate atomic data were available, we updated all oscillator strengths using the VALD3 database \citep{pisetal95,ryaetal97,kupetal99,kupetal00}. VALD3 includes the critically evaluated transition probabilities of P II \citep{miletal71,hib88}, Ca II \citep{the89}, Sc II \citep{lawdak89}, Ti II \citep{picetal02}, Cr II \citep{berlaw93,siglan90,pinetal93}, Fe II \citep{raauyl98}, Ni I \citep{wiclaw97}, and Zr II \citep{cowcor83,ljuetal06}. For the elements having more recent measurements of their oscillator strengths, we resorted to specific publications using the NIST Atomic Spectra Database \citep{kraetal13}: He I \citep{wiefuh09}, C and O \citep{wieetal96}, Mg II \citep{kelpod08a}, Si II \citep{kelpod08b}, S II \citep{podetal09}, Fe I \citep{fuhwie06,brietal91}, Y II \citep{fuhwie05}, and Ba II \citep{cur04}. The final linelist contains 90\,944 lines in the range 4490--5850\,\AA{}. The lines actually used for the abundance determinations, are collected in Table\,\ref{atomic} with their atomic data and sources. Most of the lines modeled here are weak lines formed deep in the atmospheres where LTE should prevail. They are well-suited to abundance determinations.

Although we do not expect to detect broadening due to hyperfine structure at the resolution of the GIRAFFE spectra, we did introduce hyperfine components for a few lines of Mn I and  Sc II using the atomic data for each hyperfine transition from {\sc gfhyperall.dat} \citep{kur93} in order to derive more realistic abundances for these two elements.

All damping constants were initially taken from {\sc gfall.dat} and then updated, in the event more recent and accurate values were available from VALD3. For a few Si II lines, we took the damping constants from \citet{lanetal88}.  When they are set to zeros in the linelist, the damping constants are computed in SYNSPEC48mod using the approximation given by \citet{kuravr81}. The broadening of the He I lines in SYNSPEC48mod uses specific tables, either from \citet{sha69} or \citet{dimsah84}. The broadening of the H I lines uses the tables of \citet{videtal73}.

\subsection{Determination of the atmospheric parameters}

\subsubsection{Effective temperatures and surface gravities} \label{tefflogg}

We first estimated the effective temperature and surface gravity for each star using the Geneva 7 color photometry retrieved from the WEBDA database\footnote{http://www.univie.ac.at/webda/navigation.html} and the CALIB code (North, P., private communication), based on the calibrations of \citet{kunetal97}. 
Indeed, we could not use \citet{napetal93}'s UVBYBETA routine since the 44 selected targets do not have Str\"{o}mgren's photometry. To derive \teff{} and \logg{}, we used the CALIB code which calibrates pairs of Geneva photometric indexes ($X$, $Y$), ($pG$, $pT$), and ($d$, $B_2-V_1$) 
in terms of effective temperature and surface gravity 
respectively for hot, intermediate, and cool stars. 
We  used the calibration of ($X$, $Y$) for the late B stars, ($pG$, $pT$) for the A stars and early F stars, and ($d$, $B_2-V_1$) for the late F stars.
We corrected the $B_2-V_1$  indexes affected by reddening, 
by a color excess correction E($B_2-V_1$) . As \citet{maisch84} reported that there is no differential reddening across the area covered by M6, we corrected 
$B_2-V_1$ with the average color excess $E(B_2-V_1)=0.13\pm0.01$ adopted from the literature (see Table \ref{intro}). In order to convert $E(B-V)$ or $E(b-y)$ into a Geneva color excess $E(B_2-V_1)$, we used the following two relations given by \citet{cra78} and \citet{nic81}, respectively:
$$E(b-y) = 0.74E(B-V)\,,$$
$$E(B_2-V_1) = 0.84E(B-V)\,.$$

\begin{figure}
\epsscale{.80}
\plotone{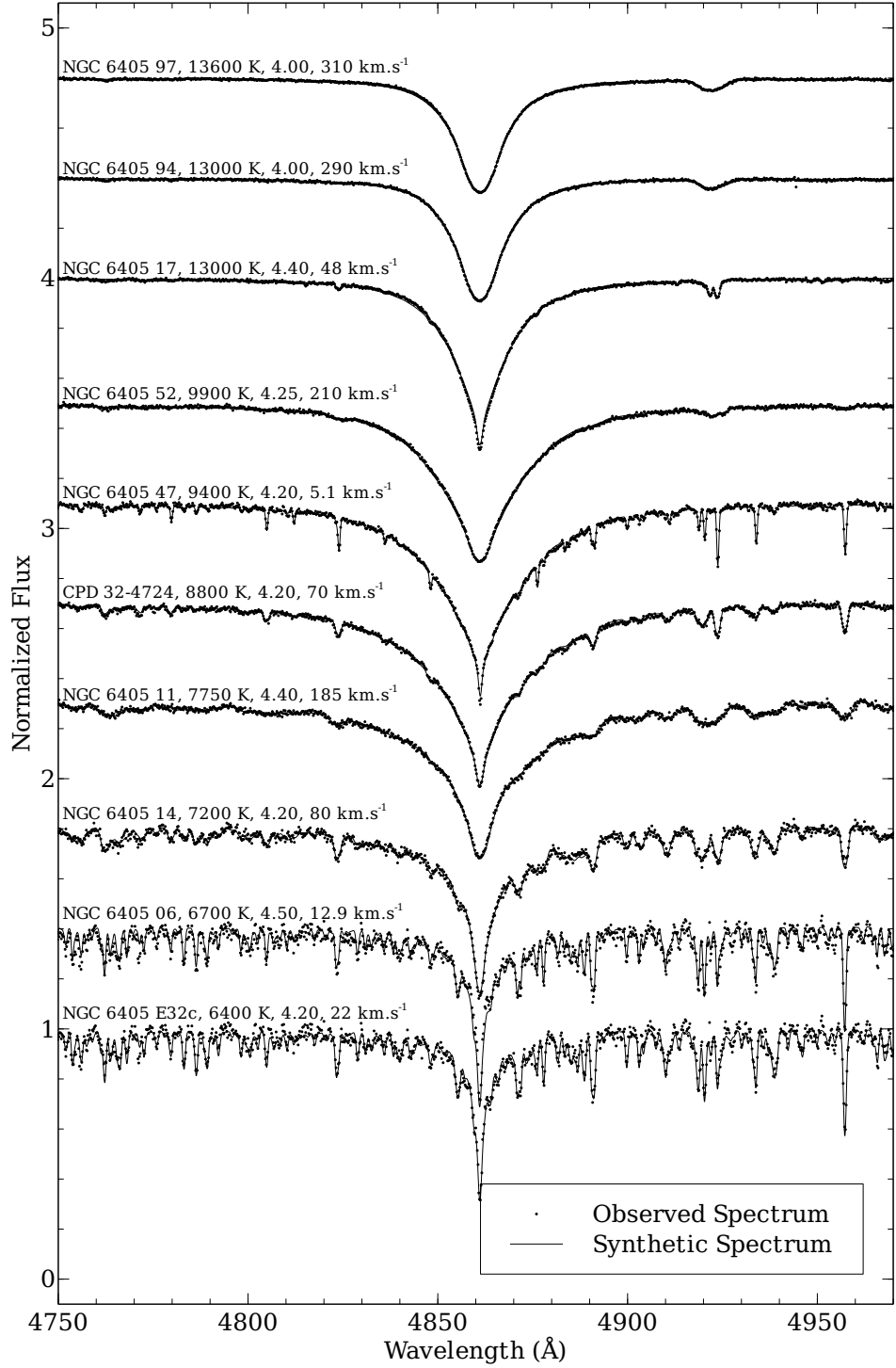}
\caption{Gallery of the $H_{\beta}$ region from the late B stars to the F stars
\label{hbeta}}
\end{figure}

We then checked these first estimates of the effective temperatures and surface gravities  i) by adjusting synthetic spectra to the observed H$_\beta$ line and ii) by studying the ionization equilibrium of Fe I and Fe II. Fig.\,\ref{hbeta} shows the comparison between the observed and best fitting synthetic H$_\beta$ line profiles for 10 stars sorted out according to decreasing effective temperature from top to bottom.
For stars with \teff{}\,$\ge$\,8500\,K, the H$_\beta$ line depends on both \teff{} and \logg{}. For this reason we also used the ionisation equilibrium of Fe (I--II)  to additionally constrain \logg{} only for the slow rotators with \vsini{}\,$\le$80\,\kms{}. We first derived the iron abundance from all unblended Fe I and Fe II lines. We then determined the set of \teff{} -- \logg{} parameters which verified the ionisation equilibrium 
(ie.\teff{} and \logg{} were altered until the iron abundance derived from the Fe I lines be equal to that derived from the Fe II lines).
We then finally checked that the found parameters did provide a good fit to the H$_\beta$ line profile. When applying the ionisation equilibrium, we considered that Fe I lines may be affected by non-LTE effects, which would lead to a slight underestimation of the Fe abundance (e.g., \citet{masetal11}). 
For stars with \teff{}\,$\le$\,8500\,K, the H$_\beta$ line profile is  sensitive primarily to variations in \teff{} and provides reliable effective temperatures. In the event the surface gravity for these cooler stars could not be derived from the  ionisation equilibrium, we adopted the photometric \logg{} value. The adopted atmospheric parameters of the analysed stars are collected in Table \ref{physical}.


\subsubsection{Projected apparent rotational velocities}

\begin{figure}
\epsscale{.80}
\plotone{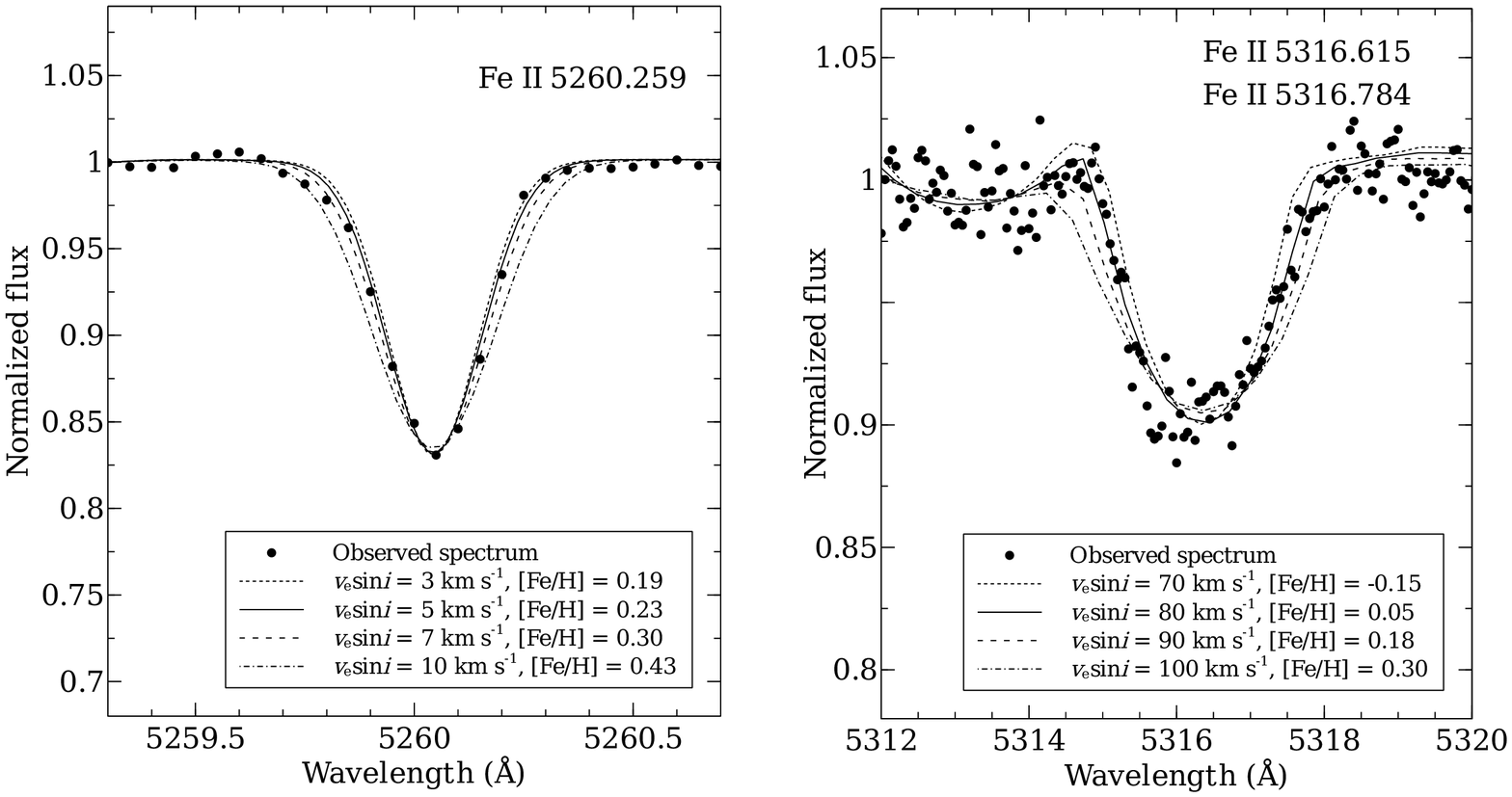}
\caption{\vsini{} determination by adjusting the synthetic spectra for the stars CD-32 13109 (left panel) and GSC07380-00766 (right panel).
\label{vsini}}
\end{figure}

The projected rotational velocities were obtained by adjusting the observed weak and moderately strong Fe II lines with synthetic spectra convolved with different rotational velocities (Fig \ref{vsini}, left panel). For faster rotating stars, with \vsini{}\,$\ge$\,80\,\kms{}, we estimated the \vsini{} value by fitting the profile of a few blended strong Fe\,{\sc ii} lines (Fig \ref{vsini}, right panel). The iron abundance was also adjusted if necessary. The final \vsini{} values were adopted by averaging the derived \vsini{} values from individual lines for each star.

\subsubsection{Microturbulent velocities}

\begin{figure}
\epsscale{.80}
\plotone{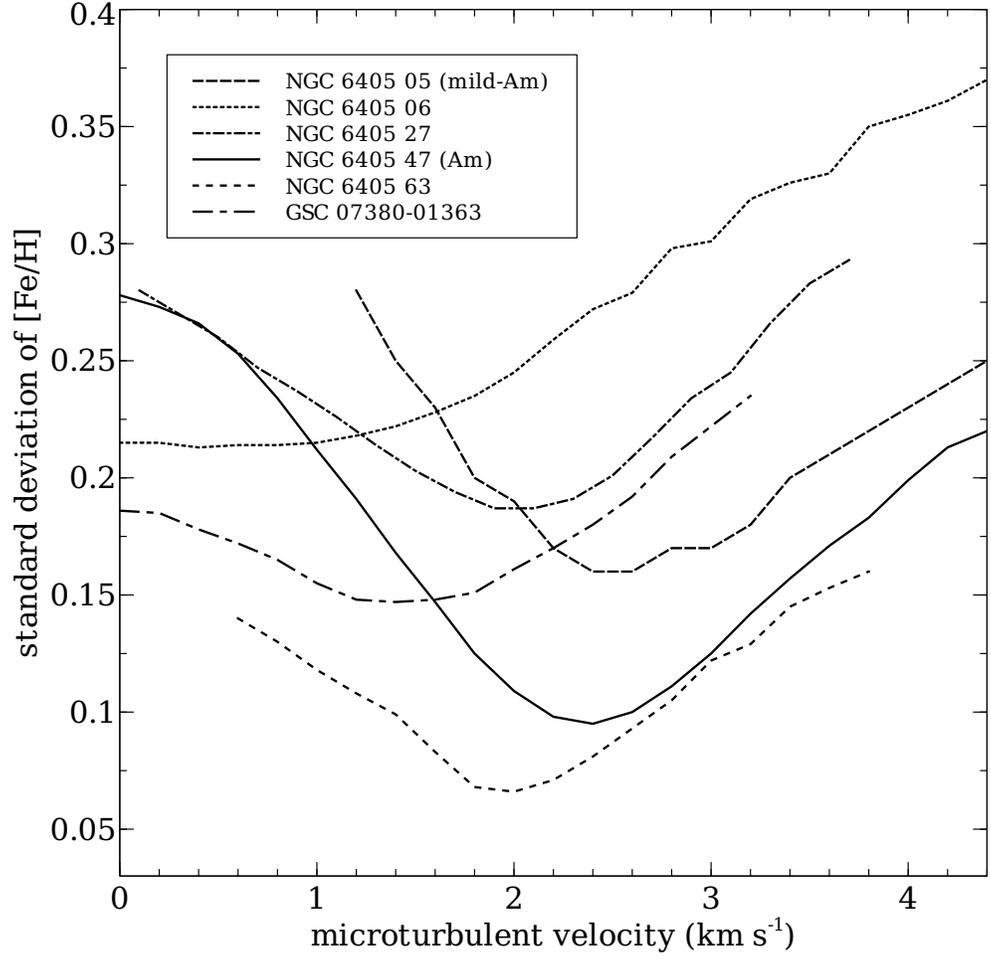}
\caption{Microturbulence determination by minimizing the standard deviation of [Fe/H].}
\label{microturbulence}
\end{figure}

To derive the microturbulent velocities, we followed the method described by \citet{fosetal07} which consists in minimizing the standard deviation of the iron abundance derived from each individual Fe\,{\sc ii} line (Fig. \ref{microturbulence}).
 In the event the number of  Fe\,{\sc ii} lines was not sufficient to apply this method (for the hottest stars), the microturbulent velocity was left as free parameter in the fitting iterations performed to derive the iron abundance (see Sect.~\ref{abundances}). We checked whether there is a difference in the derived \vmic{} using these two approaches, applying both methods to slow rotators. We found that the second approach (\vmic\ as a free parameter) typically gives only about 0.1-0.2\,km/s larger \vmic\ values\footnote{This difference mostly stems from the different line selection between the two approaches. The method of \citet{fosetal07} takes into account the unblended lines of selected singly-ionized element/elements only, while the other includes all lines of each element present in the spectrum.}. That minor difference was taken into account in the uncertainty analysis. For the fastest rotating B-type stars, very few absorption lines are available to constrain the microturbulent velocity; for this reason we adopted a null microturbulent velocity (i.e., 0.0\,\kms{}) with an assumed uncertainty of 1.0\,\kms{}, which is a value commonly reported for main sequence late B-type stars by \citet{fitmas99}. Only for one very slowly rotating chemically peculiar B-type star (HD\,318126) could we derive the microturbulent velocity, obtaining 0.5\,\kms{}.

\subsection{Abundance determinations and their systematic errors}\label{abundances}


Since the rotational velocities of the analyzed stars range from 5.1~\kms{} up to 310~\kms{},the lines can be significantly broadened by rotation.  We therefore used spectral synthesis to derive the elemental abundances. The abundances, expressed relatively to the solar ones\footnote{Recall: $\quad\left[{\mathrm{X}\over\mathrm{H}}\right] = 
\log_{10}\left({\mathrm{X}\over\mathrm{H}}\right)_{\star} - 
\log_{10}\left({\mathrm{X}\over\mathrm{H}}\right)_{\odot}$}
(\citep{gresau98})
, of twenty elements were obtained by iteratively adjusting synthetic spectra to the normalised observed spectra and by performing a $\chi^2$ minimization fit on the selected lines.
For a given chemical element, the final abundance $[{X \over H}]$
is an average of the abundances derived from each transition (i):

$$ \langle [{X \over H}] \rangle = \frac{\sum_{i} [{X \over H}]_{i}}{N} $$

where $N$ is the number of transitions for the element and $[{X \over H}]_{i}$ the abundance derived for transition (i).

For each star, the final absolute abundances (expressed as Log(X/H)) are collected in table \ref{abuntab}. As M6 is a young open cluster, we have also expressed in table \ref{cosmictab} our abundances with respect to the cosmic abundances derived for young early B stars by \citet{nieprz12}.
We expect the abundances of the following elements: Fe, Ti, Cr Mg, Mn, C, Ca and Ni to be fairly reliable as we synthesized several lines of quality A to D for these elements. For the hottest stars,  between B8 and A5, the oxygen abundance was derived mainly from the O I triplet near 5330\,\AA{}. For Y, several lines are available but their uncertainties are unknown so [Y/H] should be taken with caution.


In order to evaluate the systematic errors on the derived mean abundances, we reanalized a few representative B-, A-, and F-type stars (nos.\,17, 20, 47, 66, 99) by slightly altering the \teff{}, \logg{}, $v_\mathrm{mic}$, and the continuum level within their respective representative uncertainties: $\pm$ 150 K for \teff{}, $\pm$ 0.30 dex for \logg{}, $\pm$ 0.20 \kms  for $v_\mathrm{mic}$ and an adequate uncertainty on the continuum level depending on the line density and the apparent rotational velocity of the star analysed. The abundance was then altered until the fit to the observed profile was achieved again. This yields the abundance response to the slight change in effective temperature,
$\sigma_{\teff{}}$, surface gravity, $\sigma_{\logg{}}$, microturbulent velocity, $\sigma_{v_{\mathrm{mic}}}$ and continuum placement, $\sigma_{\mathrm{cont}}$ respectively.
Assuming these various responses are independent, we then calculated the total uncertainty on the mean abundance deduced as follows:

$$
\sigma_{\mathrm{tot}} = \sqrt{\sigma_{\teff{}}^{2} + \sigma_{\logg{}}^{2} + \sigma_{v_{\mathrm{mic}}}^{2} + \sigma_{\mathrm{cont}}^{2}}\,.
$$

The final adopted uncertainties are listed in Table \ref{Bstarerr}, \ref{Astarerr}, and \ref{Fstarerr} for B-, A-, and F-type stars, respectively.
This error analysis is similar to those performed by \citet{varmon99}, \citet{fosetal09} and \citet{sanetal13}.



\section{Results} \label{results}

\subsection{Abundance patterns}

\begin{figure}
\epsscale{.80}
\plotone{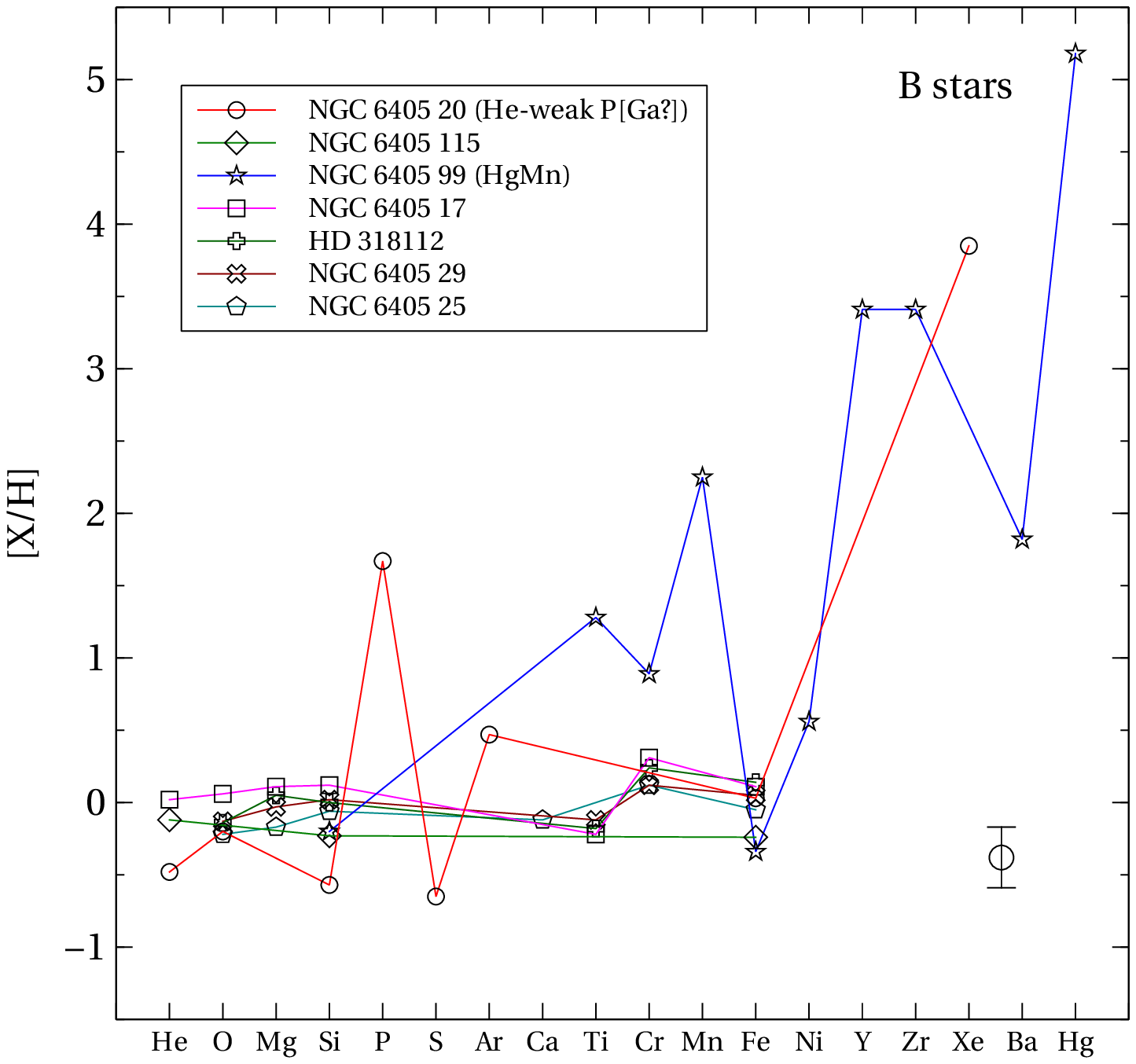}
\caption{Abundance pattern of B-type cluster members (deviations from solar abundances in \cite{gresau98}) }
\label{Bpattern}
\end{figure}

\begin{figure}
\epsscale{.80}
\plotone{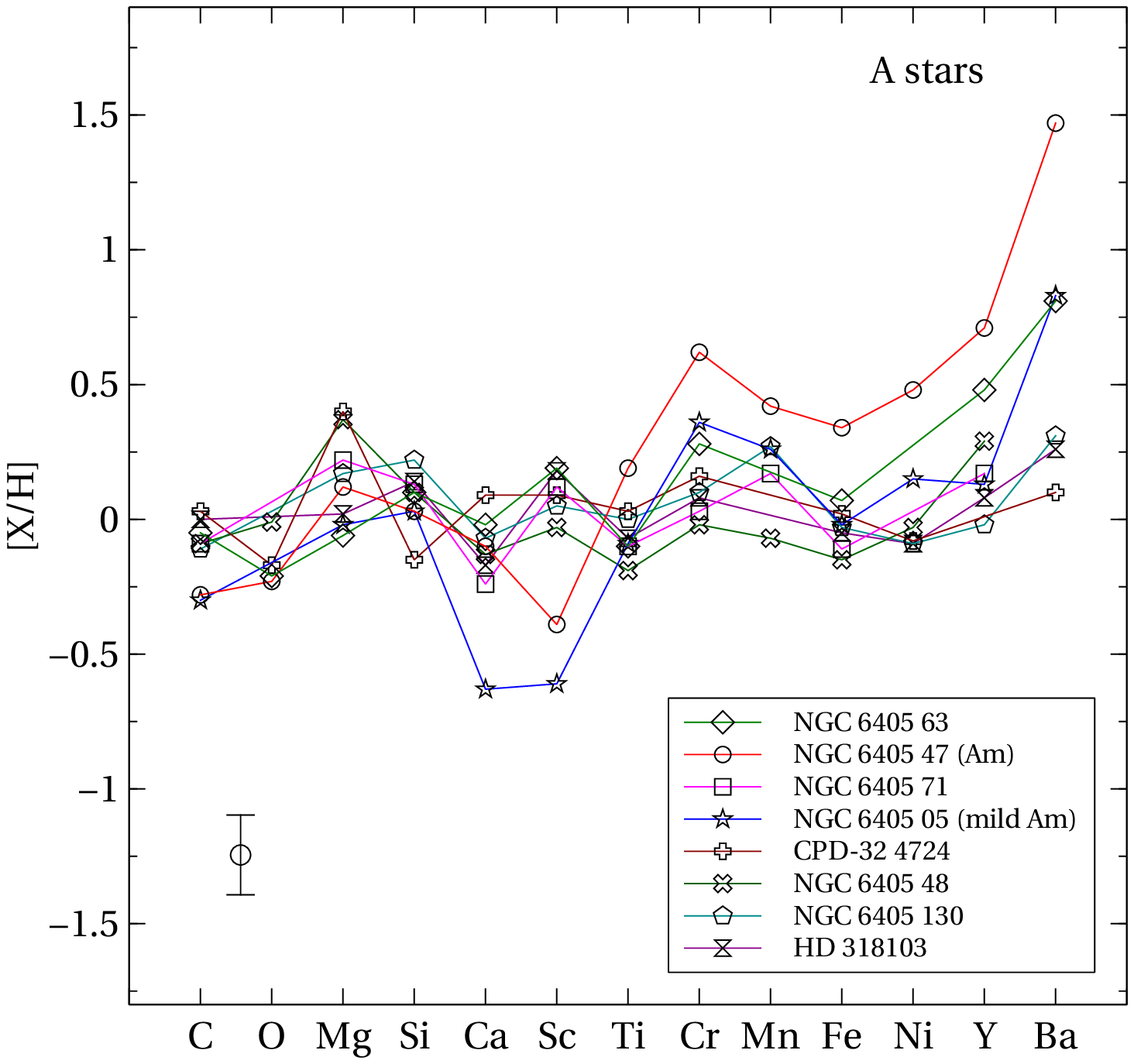}
\caption{Abundance pattern of A-type cluster members (deviations from solar abundances in \cite{gresau98})}
\label{Apattern}
\end{figure}

\begin{figure}
\epsscale{.80}
\plotone{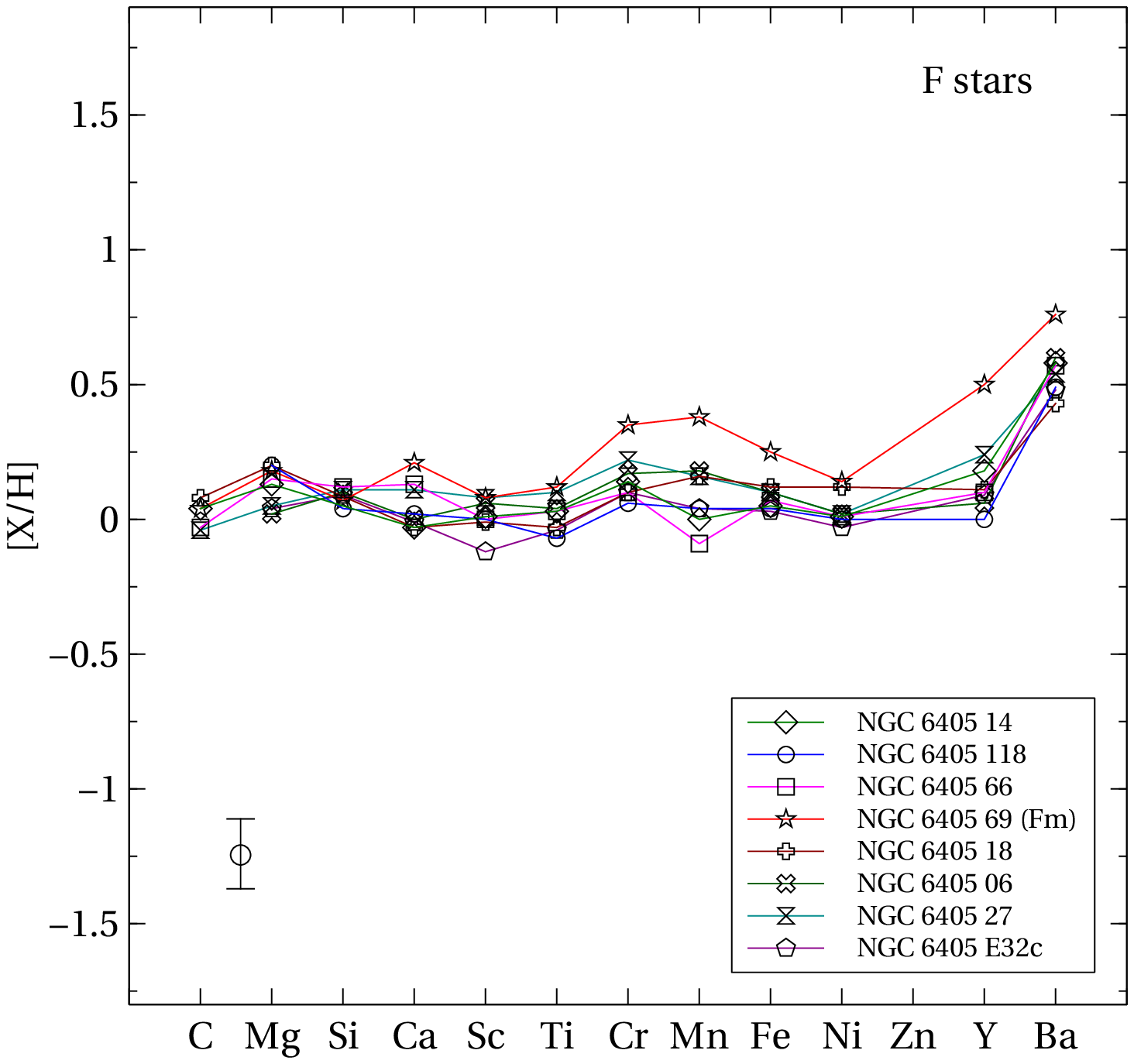}
\caption{Abundance pattern of F-type cluster members (deviations from solar abundances in \cite{gresau98})}
\label{Fpattern}
\end{figure}

Abundance pattern graphs where abundances are displayed against atomic number $Z$ are particularly useful for comparing the behaviour of the abundances of various elements for B, A and Am, F and Fm stars.
The abundance patterns of late B-, A- and F-type stars with $\vsini{}\le150\,\kms{}$ are shown in Fig \ref{Bpattern}, \ref{Apattern} and \ref{Fpattern},  respectively with representative error bars.
Inspection of these figures shows that, while the abundances of the F stars tend to be amost identical within the error bars, the abundances of the A and late B-type stars show larger and real star-to-star differences for a few chemical elements. 
The abundance patterns of the 2 Am stars (NGC 6405 47, Am) and NGC 6405 05 (mild Am) exhibit the characteristic sawtooth pattern of Am stars with a marked deficiency of scandium and overabundant elements heavier than titanium.
We stress that the existence of star-to-star variations of the abundances can be established independently of the errors in the absolute values of the oscillator strengths, since all stars will be affected in the same manner.

\subsubsection{Star-to-star abundance variations}

In order to quantify the star-to-star variation of the  abundances for a given element, we calculated the average abundance of that element and its standard deviation for the A- and F-type stars in M6 (Table \ref{startostar}). We did not perform this exercise for the  late B-type stars as most of them are  very fast rotating stars ($\vsini{}>150\,\kms{}$)  and the errors on the abundances are much larger for these stars because of the uncertainty on the continuum placement.

Among the normal A-type stars, 
the largest star-to-star variations occur for 
Ba ($\pm0.31$\,dex), Y ($\pm0.19$\,dex), Mg ($\pm0.18$\,dex) and Mn ($\pm0.17$\,dex). The star-to-star variation of [Ba/H] is much larger than the error on its abundance ($\pm0.19$\,dex), while the spreads of [Y/H], [Mg/H] and [Mn/H]  are slightly larger than the maximum abundance uncertainties which we derived for fastest rotating stars for these elements. 
We therefore believe that the star-to-star variations in the abundances of these elements are real. The abundances of Oxygen, Silicon and Calcium exhibit smaller star-to-star variations ($\sim\pm0.1$\,dex), which still remain larger than the uncertainties of their abundances.

For the normal F-type stars, the star-to-star abundance variations of the elements are mostly less than $\pm0.1$\,dex, which is smaller than  the typical uncertainties of the derived abundances. Only the Fm type star no.\,69 exhibits a slightly different pattern than the other F stars in particular differences in the Calcium, Chromium, Manganese, Iron, Yttrium and Barium abundances. This stems from the Fm character of this star. We thus fail to find any clear evidence for star-to-star variation of the abundances  for normal F stars, while the normal B and A stars tend to display larger star-to-star variations than F stars.

\subsubsection{The average metallicity of the cluster} \label{metallicity}

The F-type main sequence stars, which have substantial convective envelopes, have mixed outer layers. Their abundance patterns should thus reflect the initial chemical composition of M6. 

For each chemical element, the derived mean abundances  for the chemically-normal F-type stars of M6 are collected in column\,8 of Table \ref{startostar}, together with their standard deviations.  These mean abundances are not significantly different from solar abundances, except for Ba which is found to be overabundant (+0.53\,dex). Magnesium,  Si, Cr and Fe exhibit only marginal overabundances, which do not exceed 1.4 times of solar abundances. From this, we infer that the intial composition of M6 was nearly solar.

The mean iron abundance of M6 is found to be [Fe/H]=$0.07\pm0.03$ from the individual iron abundances of the seven normal F stars. 
The mean iron abundance can be used to derive the metallicity $Z$ of M6 using the following relation (see \citealt{fosetal11}):

$$Z_{\mathrm{cluster}}=Z_{\odot} 10^{(\mathrm{[Fe/H]}_{\mathrm{F stars}})}\ .$$

Using a solar metallicity of $Z_{\odot}=0.017$ \citep{gresau98},
the derived $Z=0.022\pm0.001$ for M6 is only marginally larger than the solar metallicity. It is also slighly larger than the present day metallicity, 0.014 $\pm$ 0.002, derived from young early type B stars in the solar vicinity by Nieves \& Przybilla (2012). This suggests that M6 formed in an environment  slightly enriched in metals compared to the current solar neighbourhoud.

\subsubsection{Behavior of the abundances with stellar parameters}

As we did previously for the other open clusters,
we have investigated the behavior of the elemental abundances against effective temperature, projected rotational velocity and the iron abundance. Any correlations/anticorrelations between the abundances and these atmospheric parameters will provide valuable constraints to theorists investigating the various hydrodynamical mechanisms affecting photospheric abundances. The abundances of the iron-peak elements, in particular manganese, should increase with effective temperature if radiative diffusion is efficient. Up to a rotational velocity of about 120 $\kms$, the abundances should not depend on $\vsini$ and may decrease at higher velocities \citep{charbonneaumi91}.

\begin{figure}
\epsscale{.50}
\plotone{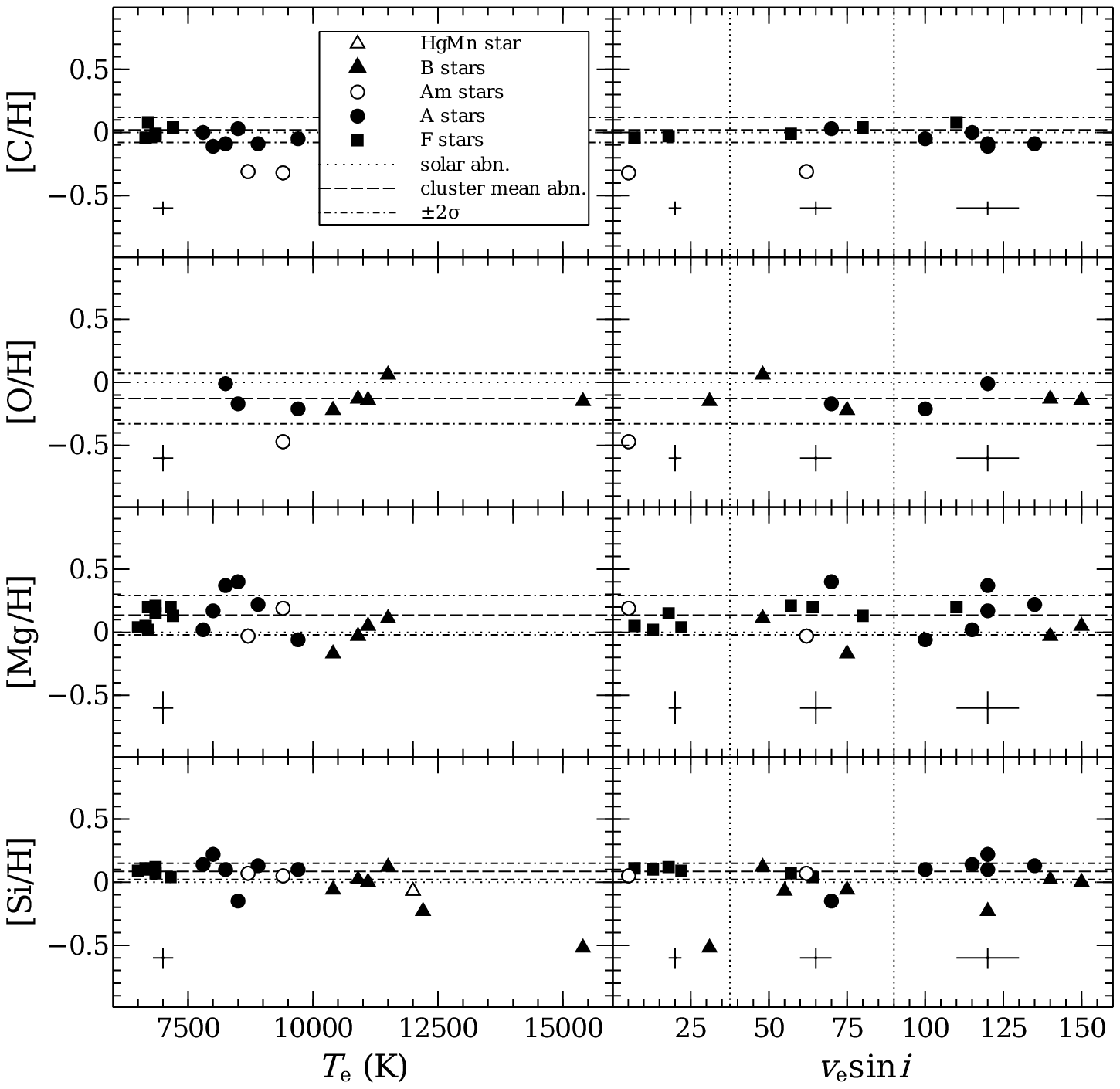}
\caption{Abundances of C, O, Mg, and Si relative to the Sun \citep{gresau98} versus \teff{} and \vsini{}.}
\label{fig:c_to_si_teff}
\end{figure}

\begin{figure}
\epsscale{.50}
\plotone{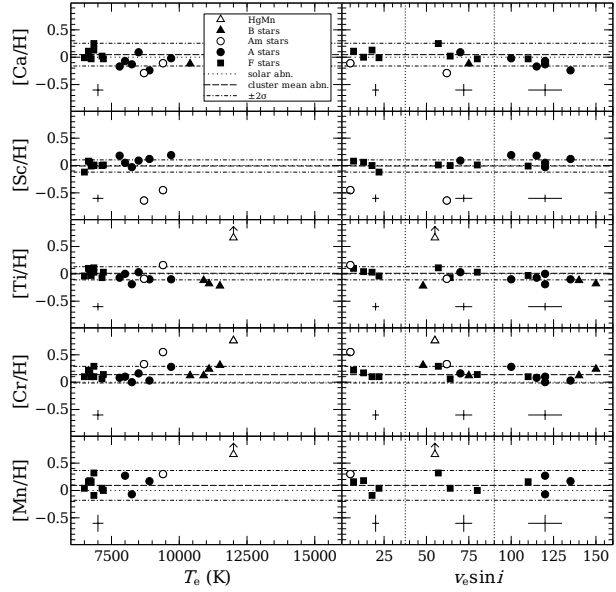}
\caption{Same as in Fig.\,\protect\ref{fig:c_to_si_teff} but for Ca, Sc, Cr, Ti, and Mn.}
\label{fig:ca_to_mn_teff}
\end{figure}

\begin{figure}
\epsscale{.50}
\plotone{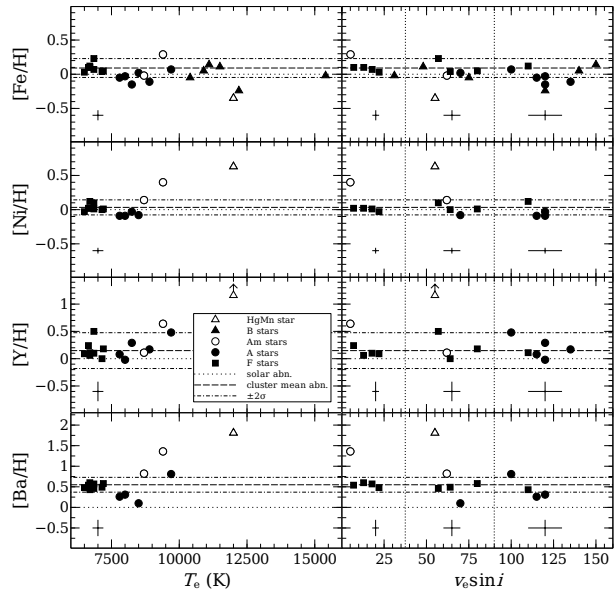}
\caption{Same as in Fig.\,\protect\ref{fig:c_to_si_teff} but for Fe, Ni, Ba, Y.}
\label{fig:fe_to_y_teff}
\end{figure}

Figs.\,\ref{fig:c_to_si_teff}, \ref{fig:ca_to_mn_teff} and \ref{fig:fe_to_y_teff} display the  abundances of C, O, Mg, Si, Ca, Ti, Cr, Mn, Ni, Sr, Y and Ba as a function of \teff{} with representative error bars. The profiles of the abundances of C, O, Mg, Ca, Ti, Cr , Mn and Sr are flat within $\pm$ 2 $\sigma$ of the mean abundance of the element, ie. their abundances are not correlated to the effective temperature. For the  normal stars, [Si/H] slightly decreases for the hottest stars only (based on 3 data only). The [O/H] and [Mn/H] profiles are flat but exhibit slightly more scatter than the other elements. One should note that  [Ni/H], [Y/H], and [Ba/H]  increase with \teff{} at $\teff{}>8500$\,K, but these trends are based on 3 data only. 

No correlations either were found between the abundances and \vsini{}. All abundance profiles against \vsini{} are flat within $\pm$ 2 $\sigma$ of the mean abundance of the element, with more or less scatter. This behaviour is expected as simulations of the abundance patterns in presence of rotation showed that abundance should not depend on rotational velocities at least up to 120 \kms since the timescale of diffusion are much shorter than those of rotational mixing \citep{charbonneaumi91}.

\begin{figure}
\epsscale{.50}
\plotone{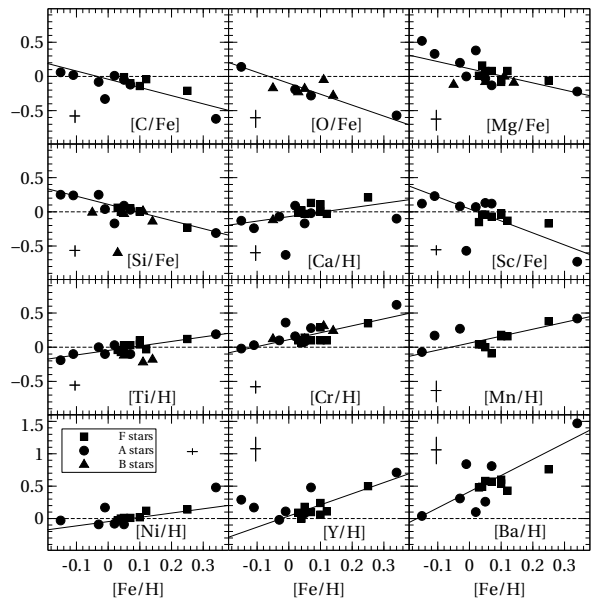}
\caption{Abundances of the elements relative to the Sun \citep{gresau98} versus iron abundance.
\label{allfe}}
\end{figure}

The abundance ratios [Ti/Fe], [Cr/Fe], [Mn/Fe], [Ni/Fe], [Y/Fe], and [Ba/Fe] appear to be correlated with the abundance of Fe, while those of C, O, Mg, Si, and Sc are anticorrelated with [Fe/H] (see Fig.\,\ref{allfe}). Table \ref{coeff} collects the coefficients of the regression analysis. Note that the abundances of magnesium and calcium do not show any convincing correlation nor anticorrelation against  [Fe/H].

\section{Discussion} \label{discussion}

\subsection{Comparison with previous studies}

In this section, we compare the abundance patterns found for M6 with those found for the other open clusters: Praesepe, NGC 5460, Coma Berenices,  the Pleiades and the Hyades analysed by \citet{fosetal07}, \citet{fosetal11}, \citet{gebetal08}, \citet{gebmon08}, \citet{gebetal10}, respectively.

The analysis of these clusters revealed a common trend: the star-to-star variations of the abundances of most chemical elements are usually larger for A stars than for F stars. In their analysis of Coma Berenices (age about 450 Myr), \citet{gebetal08}, found that the A stars exhibit larger star-to-star variations for C, O, Na, Sc, Ti, Mn, Fe, Ni, Sr, Y, Zr, and Ba, than the F stars do. 
For the Pleiades (age about 100 Myr),
\citet{gebmon08} showed also that A stars exhibit larger variations in C, Sc, Ti, V, Cr, Mn, Sr, Y, Zr and Ba, than the F stars. 
For the Hyades (age about 700 Myr), \citet{gebetal10} found significant star-to-star variations of the abundances of C, Na, Sc, Fe, Ni, Sr, Y, and Ba for the A stars
(ie. the maximum abundance spread is larger than three times $\sigma$ ,the representative uncertainty on the mean abundance of the considered element).
Our findings for the M6 open cluster (age about 75\,Myr) similarly confirm that  normal A stars exhibit a larger scatter of their abundances around the cluster for Mg, Si, Ca, Cr, Mn, Fe, Y, Ba, than the F stars do. We note that the  Cr, Mn, Fe, Y, and Ba abundances exhibit significant star-to-star variations among the normal A stars for each cluster.  We believe that these variations can be interpreted as evidence of transport processes induced by the larger rotation rates in these stars and competing with radiatively driven diffusion. 

Similarly, the anticorrelations of [C/Fe], [O/Fe], [Mg/H], [Si/H] and [Sc/H] with [Fe/H] and the correlations of [Ti/H], [Cr/H], [Mn/H], [Ni/H], [Y/H] and [Ba/H] with [Fe/H] in M6 can be compared with those found by \citet{gebetal08} and \citet{gebetal10} for the Coma Berenices and Hyades cluster in terms of both slope and sign of the regression. 
\citet{gebetal08} found correlations between [Mn/H], [Ni/H], [Sr/H], [Ba/H] versus [Fe/H] and anti-correlations for [C/Fe] and [O/Fe], for the A and Am stars in Coma Berenices. Our findings for M6 strongly support these findings: [C/Fe] and [O/Fe] exhibit anti-correlations and most other elements exhibit strong correlations with [Fe/H] as predicted by models including diffusion.
In particular, we note that the slope of the anticorrelation of [C/Fe] vs. [Fe/H] in Fig \ref{allfe}, which equals to -1.20 for M6, is only slightly larger than -1.74 found by \citet{gebetal08} for 15 A and F stars in Coma Berenices  and  agrees well with -1.24 found by \citet{hil95} for 15 sharp-lined field A stars. 

\subsection{Comparison to the predictions of self-consistent evolutionary models}

\subsubsection{The Montr\'eal OPAL-based stellar evolution code}

The derived abundances have been compared to the predictions of recent evolutionary models computed 
with XEVOL, the Montr\'eal stellar evolution code\footnote{From the Georges Michaud astrophysics group.}.  
This code evolved from a lagrangian stellar evolution code written by D. A. VandenBerg and 
the Victoria Group (\cite{pedersenetal91}).  \citet{proffittmi91a,proffittmi91b} added an accurate, 
implicit treatment of atomic diffusion (including thermal diffusion) and 
gravitational settling of all major isotopes of H, He, C, N, and O, 
and some trace elements (Li, Be, B);  various simple expressions were used to compute radiative forces
on a few test elements.   They also introduced and explored the effects of a simple density 
power-law model of \emph{turbulent diffusion}, as a means of smoothing composition discontinuities 
near convection zone boundaries, in a physically reasonable and parametrizable way.  

The Victoria code relied on multi-dimensional interpolation among pre-calculated 
Rosseland opacity tables, like most stellar codes of that time.  
One of the most promising sets of opacity tables was the set developed entirely from theory 
by the OPAL project (\citealt{rogersig92}); this and many other similar alternatives were used and compared 
by different authors for many years; they all had a common weakness: they did not provide a means 
of reliably computing radiative forces \emph{on-the-fly}.  Some XEVOL experiments were made with the GLAM 
method of radiative force calculations\footnote{This method can be used for any atom whose basic 
atomic data are known.} (\citealt{glam95}) but the method is too time-consuming for stellar evolution applications.

\citet{richermiroetal98} introduced into XEVOL the new OPAL96 opacity data provided by \citet{iglesiasro96}; 
these contain not only OPAL's new Rosseland mean opacity tables, but also the full spectra of all elements 
in their mix, at all tabulated points, plus tables of the average ionic charge of each element\footnote{Average ionic 
charges are used to setup diffusion equations in terms of \emph{average ions}, for elements other than H or He.}.  
These allow the easy computation of various mean opacities and radiative accelerations at any point in the star, 
for any local composition.   The spectra of all elements (24 distinct atomic numbers are 
currently available\footnote{of which three (Li, Be, B) where computed by the Montr\'eal group, 
not by OPAL.}) are separately given for standard abundances and standard temperatures and densities 
(using $R\equiv\rho/[T/10^{6}\Kelvin]^3$ as the density variable), on a common $10^4$-point grid 
of the dimensionless frequency $u\equiv h\nu/kT$.  

Elemental spectra are first scaled according to the local abundances of the stellar model shell, 
on each point of a small $(T_i,R_j)$ sub-grid\footnote{Usually of size $2\times2$, for speed.  
To each $(T_i,R_j)$ pair correspond 24 spectra of individual elements, 
plus other plasma properties, at standard abundances.} of 
the OPAL96 tables bracketing the actual $(T,R)$ parameters of the shell; 
the spectra are then added, inverted as necessary, then integrated over frequencies; 
the resulting local sub-grid values for the function ($\kappa_\mathrm{R}$, $g_\mathrm{rad}(\mathrm{element})$, 
or other property of the plasma under consideration) are then combined using the electronic density $N_e$ 
and temperature $T$ as interpolation variables 
(a requirement of the OPAL \emph{method of corresponding states})\footnote{Integrations must be performed 
before interpolations, as explained by \citet{richermiroetal98}, because of the $T$ dependence of physical 
frequencies  corresponding to given dimensionless frequencies $u$.}, to compute the desired local 
quantity at the layer $(T,R)$; local radiative accelerations and mean opacities 
(and their thermodynamic derivatives) are calculated in this way.  This is repeated at each model layer 
and for each time step, and repeated as necessary until convergence of the model equations are 
satisfied everywhere.  Such calculations were not possible until the advent of fast and 
inexpensive parallel computers.

Radiative accelerations include corrections factors for the redistribution of momentum among ionization states from \citet{glam95} and \citet{leblancmiri2000}; these become progressively more important as one approaches the star surface, and are not computable from OPAL data.

\citet{turetal98a,turetal98b} introduced finite element methods early in the development of the Montr\'eal code, 
to improve the treatment of diffusion (element conservation in conditions of extreme concentration variations) 
and to extend it to all elements in the model; they adapted it to allow accretion and pulsation studies;
\citet{vicetal10} added the handling of \emph{mass loss}, either homogeneous or 
inhomogeneous\footnote{Major modifications to the mesh control and optimization algorithm now allow 
the evolution to proceed through the He-flash (with a simplified treatment of the flash event itself) to the first 
half of the Horizontal Branch evolution, the code current limit corresponding to the time when 
both H and ${}^4$He are completely destroyed in the star center ($X=Y=0$).}.      
The evolution code was modified to allow diffusion physics to cover the whole envelope, 
right up to the atmosphere, 
the first surface layer (and model outer boundary) being allowed to be as thin at $10^{-15}$\,M$_\odot$.   
New turbulence models and new parameterizations were introduced; the latest version of the code 
allows for multiple, uncoupled or weakly coupled turbulent zones bordering convection zones.   
This makes possible the study of some hot main sequence stars, where the outer envelope is thought 
to harbor a complex multi-zone convective structure.  


\subsubsection{Modeling of the P-rich star HD\,318101}\label{modelingHD318101}

We have specifically attempted to model the abundance pattern of HD\,318101 
(cf. Sec\,\ref{sec:NGC_6405_20}), which we identify 
as a new P-rich star in M6.     This star belongs to the He-weak/Hg-Mn region of the HR diagram 
(cf. Fig.\,1.1 and Sect.\,8.1--8.3 of \citet{MichaudAlRibook2015} for a general discussion 
of diffusion phenomena in these stellar classes).  

A proper simulation of element diffusion in this type of stars requires a detailed analysis of radiative forces and 
diffusion in the atmospheric layers of the star;  unfortunately, XEVOL is of limited help in 
\emph{stable, optically thin} atmospheric layers, which appear to be 
playing a major role in the development HD\,318101 surface anomalies;   the code only knows 
about LTE physics, the diffusion approximation of the radiative transfer problem, and cannot deal 
with non-local atmospheric radiative transfer.    Nevertheless, we assume it still gives a 
reasonable working approximation for the overall structure of the atmosphere, 
but our radiative forces and abundance profile solutions cannot be trusted there,  
as illustrated convincingly by the work of \citet{alecianstift2010} and \citet{AlecianStDo2011}, 
and summarized in \citet[\S\,8.1--2]{MichaudAlRibook2015}.  
For that reason, XEVOL has until now only been applied to cases where atmospheric layers could be  
assumed to be well mixed at all times by some turbulence generating mechanism (e.g., a nearby convection zone).  
The optically thick parts of the star should be well modeled by XEVOL.

In the case of HD\,318101, XEVOL can predict the evolution of element distributions in the stellar 
core and envelope, and give, if so desired, quantitative information about the initial \emph{flux} of each element into  atmosphere, through the atmosphere/envelope boundary. 

Once the evolution of the inner parts are sufficiently well characterized, the atmospheric layers could 
be handled in a way similar to the approach of \citet{moehlerdrleetal14} or \citet{alecianstift2010}, 
to obtain a complete solution (some iterations of the whole process may be necessary for fine tuning 
of $M_*$, $L_*$ and $\teff{}_*$ at the desired ages).   
The simulation data shown below must be considered as a first approximation to the solution.

The mixing length parameter and the initial He abundance for our HD\,318101 models
($\alpha = 2.096$ and $Y_{0} = 0.279$ respectively) were calibrated 
to fit the current luminosity and radius of the Sun (see \citealt{turetal98b}, model~H). 
The models follow the evolution of all its elements, including a few of their isotopes 
(${}^2$H, ${}^3$He, ${}^6$Li, ${}^7$Li, ${}^9$Be, ${}^{10}$B, ${}^{11}$B, ${}^{13}$C), up to nickel. 
Models were evolved from the pre-main sequence with a solar scaled metals mixture 
(these initial abundances are listed in Table\,1 of \citealt{turetal98b}). 
The initial (zero age) mass fraction of metals was set to $Z_0 = 0.02$\,.
In models with little or no mass loss, the atmospheric layers must be kept homogeneous to prevent 
excessive buildup of some elements at the surface, as a result of their large outward fluxes, and because
of the limitations of our atmosphere modeling as we have just explained above;  
this was done numerically from the surface down to a temperature of 65\,000\,K, at every time step;
the interior limit corresponds to the He\,\textsc{ii} convection zone.

The self-consistent modeling proceeds as described in \citet{ricetal00};  turbulence is needed 
to prevent excessive anomalies to develop very rapidly even in the envelope.    
Various profiles have been used in a search for conditions that would allow phosphorus and 
other elements to rise to the  surface, while maintaining a near normal iron abundance at the age of the star; 
we allowed in our search an age window of  50--100\,Myr.   A temporary rise in Fe would be acceptable as long as 
it returned near solar values at the time of the observations.

In the \citet{ricetal00} study, only one surface ``convection'' zone was considered: a zone including 
all mass from the surface down to the inner boundary of the deepest surface zone 
(it could be the He\,\textsc{ii} or the Fe convection zone inner boundary).  
To the assumed complete mixing in this large convection zone, turbulence below the zone was added 
in the form of a turbulent diffusion coefficient, $D_\mathrm{T}$, according to various profiles 
and only down to layers where atomic diffusion becomes dominant over turbulent diffusion.
The cause of this turbulence was not considered.\footnote{It was found to be compatible with 
rotationally induced turbulence as calculated by \citet{zahn2005} and \citet{talonrimi2006}.}   
In these conditions, it was shown that the resulting anomalies only depend on the initial metallicity 
plus one extra parameter that was called the ``effective mixed mass'', not on the detailed shape 
of the turbulence profile.  Fig.\,12 of that article shows abundance patterns for a star of 
2.5\,M$_\odot$ starting with $Z_0=0.02$; the models all share a similar P overabundance typically 
in the +0.3 to +0.7\,dex range.   Their Fig.\,14 shows that [Fe/H] tends to saturate at about +0.3\,dex 
(taking into account the rise in H), while the P rise does not saturate; it follows Fe at 100\,Myr, 
but continues to rise as evolution continues; Fig.\,22 shows one case where the P abundance 
anomaly is about twice that of Fe; but this is in an older and cooler star ($\teff{}=7500$ at 800\,Myr).

For the calculations presented here, we considered stellar masses in the range 
4.5--5.0\,M$_\odot$.   Surface homogenization was always applied between the surface and 
the layer at which $T=65\,000$\,K, to speed up calculations and avoid instabilities in the optically thin regions.  
This amounts to assuming that the He\,\textsc{ii} convection zone is turbulent enough to eventually mix also 
all the  layers above it in a short time, compared to the lifetime of the star.

\begin{figure}
\epsscale{1.0}
\plotone{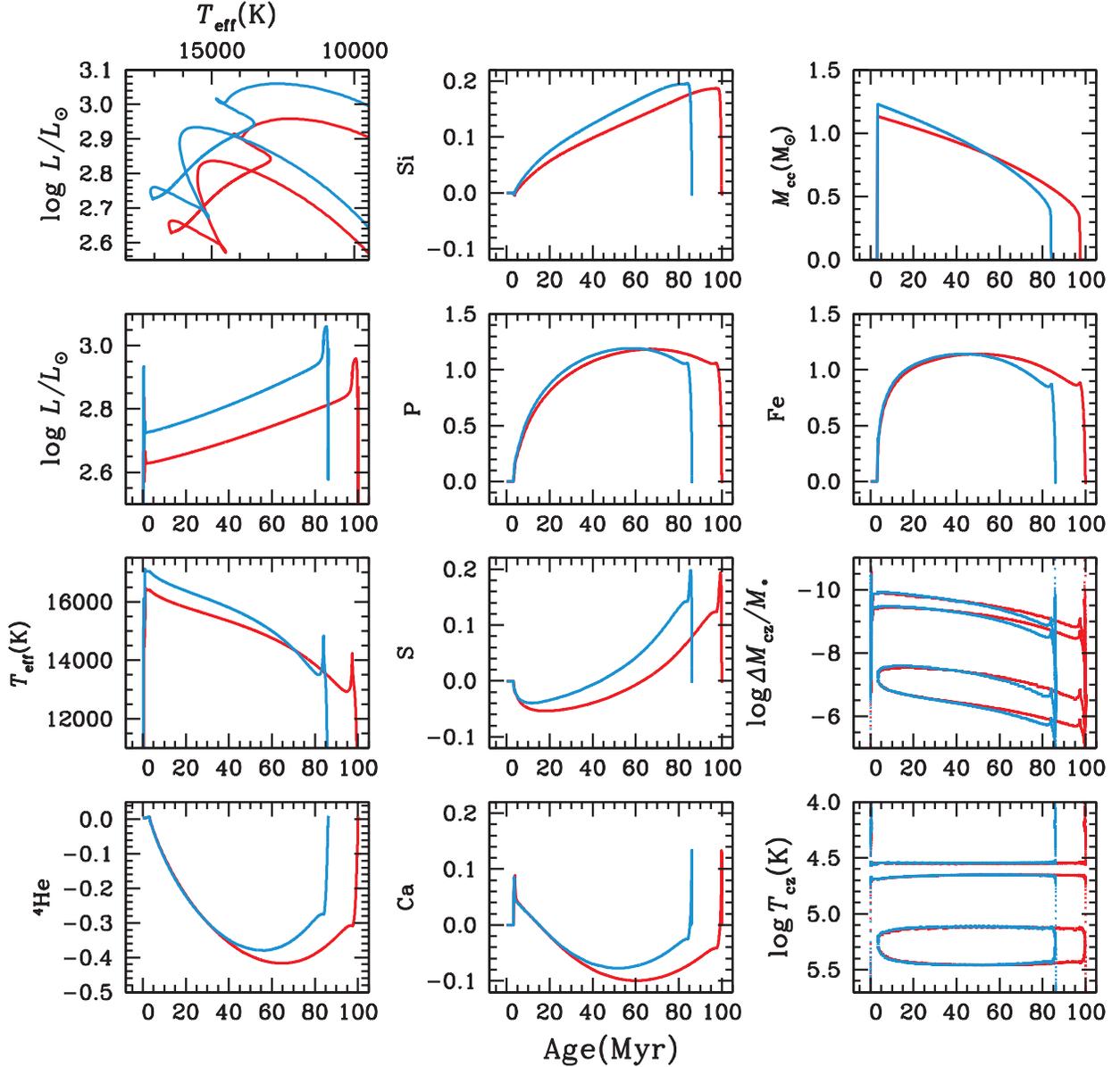}
\caption{Evolution of properties of two models for HD\,318101, 
with $M_*=4.7$  (red) and $5.0\protect\Msol$ (blue).  The initial metallicity $Z_0=0.02$.
Surface layers are homogenized down to the bottom of the He\,\textsc{ii} convection zone.  Extra turbulence 
\texttt{T150KD1K-3} is added (see text), covering mostly the interior of the Mn-Fe-Ni convection zone; the red curves correspond to the T150K model of figure~\protect\ref{M6_vs_xevol1}.  $\Delta M_{\mathrm{cz}}$ is mass measured from the star surface.  Panels for chemical elements show mass fraction changes $\log_{10} (X/X_0)_{\mathrm{element}}$ at the surface.
\label{melody_star_internals}}
\end{figure}

Fig.\,\ref{melody_star_internals} presents the evolution of some internal properties of these models 
for HD\,318101, with masses 4.7\,M$_\odot$  (red curves) and 5.0\,M$_\odot$ (blue curves), 
which should be representative of that star.   The Mn-Fe-Ni convection zone was allowed to develop 
on its own\footnote{which it did in about 3.5\,Myr.}, spreading between $\sim$130\,000\,K and $\sim$300\,000\,K, 
but some mild turbulence was added there, to help manage the proliferation of tiny convection zones in that area, 
by helping them fuse.   The turbulent diffusion coefficients used where of the form (see \citealt[\S\,7.3.3]{MichaudAlRibook2015})
\begin{equation}
D_\mathrm{T}(\rho) =    \omega D_\mathrm{He}(T_0)\  (\rho/\rho(T_0))^{-3}\,,
\end{equation}
where $\rho$ is the local density, $D_\mathrm{He}(T_0)$ is the diffusion coefficient of ${}^4$He in 
hydrogen at $T=T_0$ in the model, and $\omega$ is a constant controlling the overall strength of the profile, 
relative to atomic diffusion of He.\footnote{\protect\citet{huetal11} used a similar expression for $D_\mathrm{T}$ in the context of sdB stars.}  
The reference $T_0$ was set to 150\,000\,K and the turbulence 
scale $\omega$ was set to 1000 in both cases.\footnote{According to past XEVOL model naming conventions, 
these diffusion models would be called: \texttt{M4.7000DiffT150KD1K-3} and \texttt{M5.0000DiffT150KD1K-3}, 
where the letters  \texttt{M,T,D} are followed by the star mass, 
the turbulence profile anchoring temperature $T_0$, and the scale 
$\omega$ and slope of the turbulence profile; often, $\log T_0$ is used instead of $T_0$ after \texttt{T}.  
Mass loss (stellar wind) is labelled \texttt{W}, when present.  This convention is also adopted in the present article; we will use shortened names whenever context allows it.}

In this short mass range, most red and blue curves 
look very much like time-stretched copies of each other.   The effective temperatures 
would be compatible with the observed one (within error bars) for an age in the range of 50--70\,My.
Panels for elements He to Fe show surface concentration changes  $\log_{10} (X/X_0)_{\mathrm{element}}$.   
The convection zone (cz) panels show the depths of all the convection zone boundaries in the atmosphere and the envelope, in the course of time.  $M_{\mathrm{cc}}$ is the convective core size, in solar masses.

P and Fe follow each other closely, as in the less massive \citet{ricetal00} models mentioned above, but with P climbing slightly more than Fe.  The observable abundances (taking atmosphere physics into account) would depend on how these large overabundances redistribute themselves within the atmosphere, in a locally stratified structure;   some Fe may actually be ejected from the surface while P, having fewer supporting lines, may remain trapped there, leading to the observed pattern in HD\,318101.



\begin{figure}
\epsscale{1.00}
\plotone{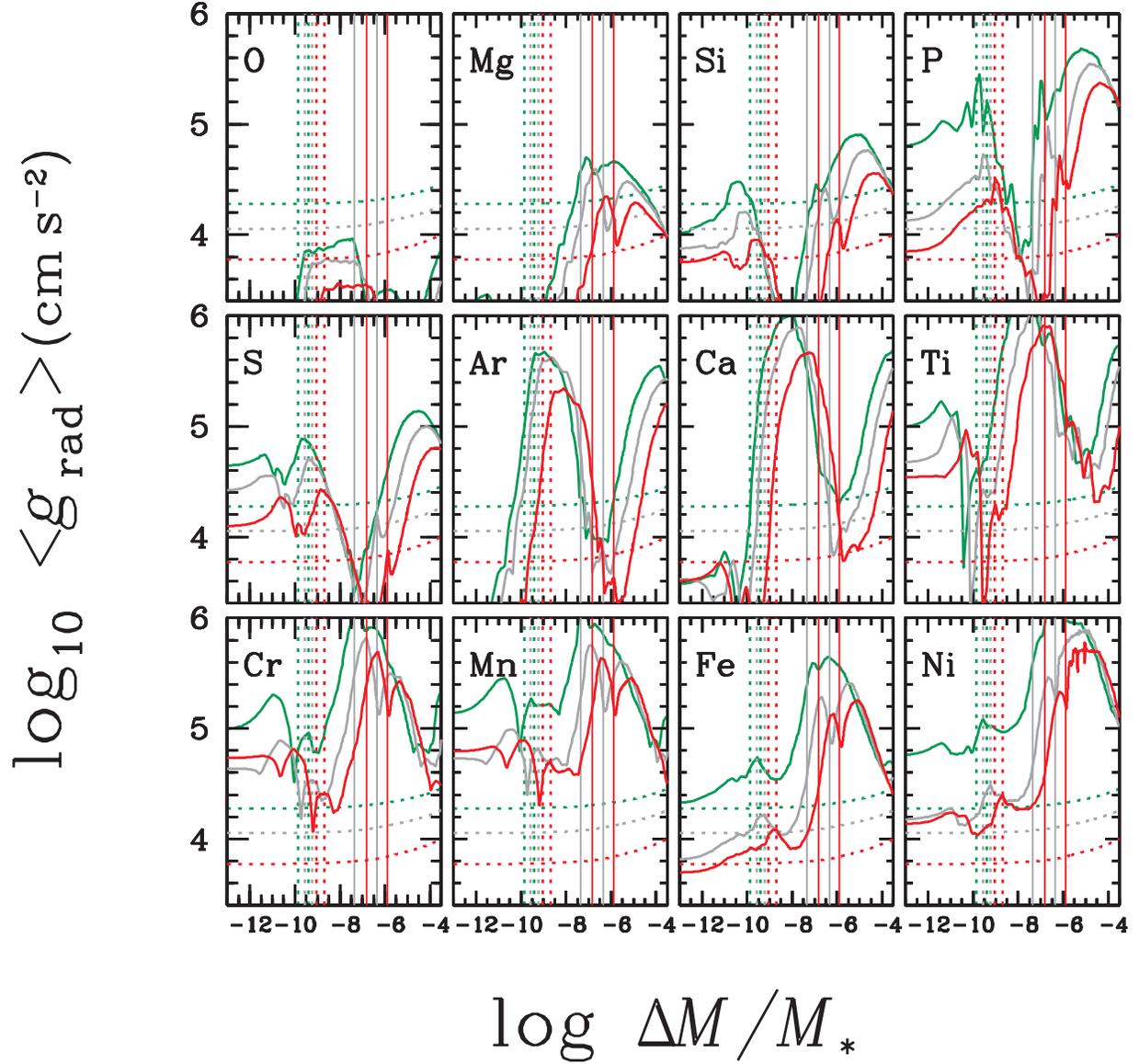}
\caption{Radiative acceleration profiles from XEVOL for a number of elements in the $4.7\protect\Msol$ model 
\texttt{T150KD1K-3}.   Profiles are shown for ages 2\,Myr (green), 50\,Myr (grey), and 85\,Myr (red).  
The horizontal dotted curves show the run of
$\log g$ through the envelope, while vertical lines show the limits of envelope convection zones, for  
He\,\textsc{ii} (dotted) and iron-group elements (solid, if present).
\label{M6_gradM4.7T150K}}
\end{figure}

Fig.\,\ref{M6_gradM4.7T150K} shows the $g_\mathrm{rad}$  profiles of a number of elements 
in the 4.7\,M$_\odot$  \texttt{T150KD1K-3}   model, at what are roughly the beginning (50\,Myr, grey), 
and the end (85\,Myr, red) of the age window  
considered acceptable for HD\,318101.   Profiles are also shown prior to the appearance of the Mn-Fe-Ni convection zone,
at age 2\,Myr (green curves).
The evolution towards the right of all the curves is the result of the cooling
of envelope layers during that period. The limits of the He\,\textsc{ii} and Mn-Fe-Ni convection zones are indicated by
vertical bars (\emph{dotted} for He and \emph{solid} for Mn-Fe-Ni).   One notices a marked difference between  P  and Fe 
behaviour in  the vicinity of the He convection zone, with P being more strongly pushed towards the surface in that region;
P also appears to be wanting to leave the star, while Fe does not; but as explained above, 
the atmospheric part of the model is
not sufficiently reliable to draw such conclusions, especially in the present case 
where $g_\mathrm{rad}(\mathrm{P})\approx g_\mathrm{rad}(\mathrm{Fe}) \approx g$ in these layers after 50\,Myr.   
The weak $g_\mathrm{rad}(\mathrm{Fe})$ in the upper envelope is in good part the result 
of lines saturation; we see that Cr, Mn, and Ni are less affected by it, due to their lower abundance.  
As time passes, a given peak of the deeper force profile of an element pushes that element up 
with a decreasing force, into an increasing amount of mass.  
Elements P and Fe are both pushed into the Mn-Fe-Ni convection zone from below; 
P should accumulate there, but not Fe, which continues to rise towards the surface.  
Phosphorus within the topmost $\sim\!10^{-8}\,M_*$ of the envelope should also rise towards the surface, 
potentially allowing overabundances in excess of a factor $10^3$ to $10^5$, 
limited by a growing opposite diffusion gradient in the atmosphere 
(or by mass loss, if there were some); it is line saturation which prevents Fe from growing 
even more than P at the surface.

Fig.\,\ref{M6_gradM4.7T150K} suggests that one might be able to reproduce the P and Fe abundances of HD\,318101
(high P, normal Fe) if one can avoid mixing surface layers down to the bottom of the  He\,\textsc{ii} convection zone, and stop evolution at an early age (e.g., $\leq50$\,Myr).  This would require extremely short time steps, and keeping the Mn-Fe-Ni convection zone well mixed to prevent that region from taking control over the evolution.

\subsubsubsection{Convection zones}

\begin{figure}
\epsscale{0.68}
\plotone{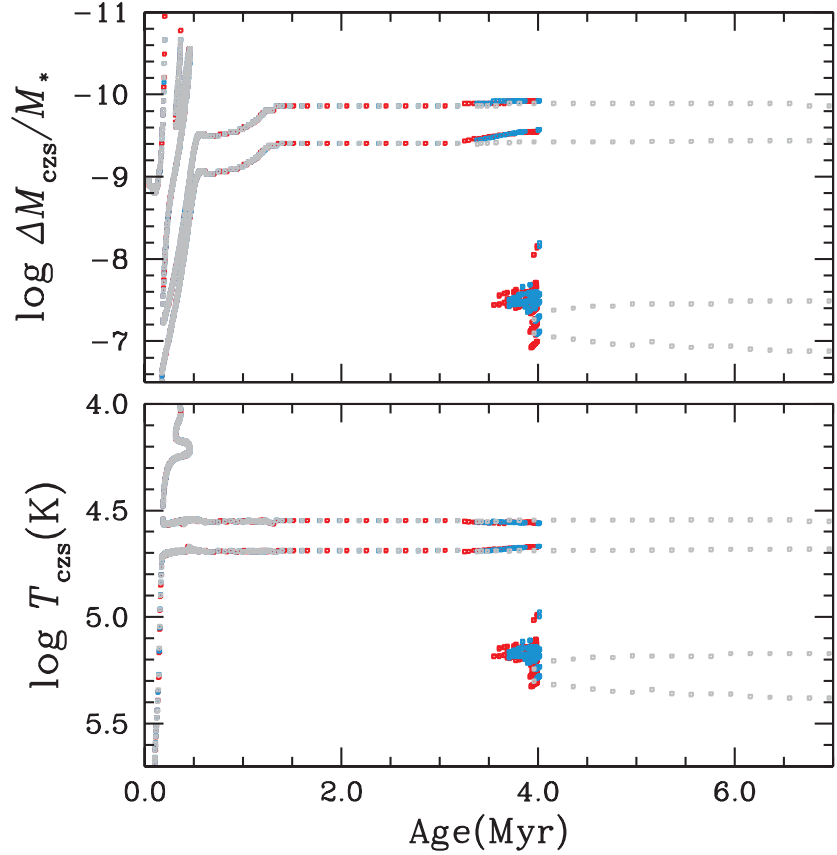}
\caption{Envelope convection zone boundaries in an evolving 4.7\protect\Msol{} star with initial metallicity $Z_0=0.02$ and the same initial \emph{metal proportions} as in the best XEVOL age zero solar model.  
The figure shows the boundaries for every converged model in the time sequences.  
Three models are shown (see text for details): 
\texttt{T65KD10-4} (blue), with
weak, short range turbulence below the He\,\textsc{ii} convection zone; 
\texttt{T150KD1K-3} (grey), with 
much stronger turbulence in the iron convection zone only (grey);  
\texttt{T65KD30-4W5E-15} (red), with turbulence 3 times the strength of that in the blue model, 
plus weak, constant stellar mass loss ($5\times 10^{-15}\Msol$/yr).   Homogeneity is artificially enforced in layers cooler than 65\,000\,K ($\log T \lessapprox 4.8$).
\label{4.7Msol_zc}}
\end{figure}

\citet{ricetal01} looked at convection zones caused by iron-group elements in main-sequence 
stars of various metallicities ranging from $Z_0=0.01$ to 0.03. Their Fig.\,8c shows the full 
evolution of  convection zones in a 2.5\,M$_\odot$ star.
The iron convection zone (labelled $\gamma$ in their figure) is getting wider with increasing stellar mass.
The situation in HD\,318101 follows that trend, but appears to be more complex, probably due to its 
higher mass and effective temperature, and shorter diffusion timescale (\citealt{ricetal01,richard2005}). 
The most important obstacle to simulation is the chaotic convective structure that develops when Fe, Mn, and Ni 
become abundant enough to trigger convection; these elements are subject to radiative force saturation, 
which limits their overabundances in ``iron-group'' convection zones, but also leads to their spreading 
(as shown by \citealt{ricetal01}); this is illustrated here in Fig.\,\ref{4.7Msol_zc} for tentative models for 
star HD\,318101.  The detailed structure of these convection zones, whether they lump together or split into many parts, depends sensitively on the shapes of the rapidly changing Fe, Mn, and Ni abundance profiles.  Numerical noise in these profiles can contribute to the convection zone(s) fragmentation; some of that noise originates in the bilinear 
$(R, T)$ interpolation method used with OPAL grid data.
Some localized turbulence is introduced to help smooth out that evolution.

Three simulations are shown.  The first one is \texttt{T65KD10-4} (blue), with weak, short range turbulence 
limited to the hot boundary of the He\,\textsc{ii} convection zone;  model \texttt{T150KD1K-3} (grey) 
has much stronger turbulence effectively limited to the Mn-Fe-Ni convection zone only (grey); finally, 
model \texttt{T65KD30-4W5E-15} (red) has turbulence 3 times the strength of that in the blue model, 
plus weak, constant stellar mass loss ($5\times 10^{-15}\Msol$/yr) introduced to peel off the topmost layers 
of the star, where excessive accumulation of some elements may occur.   
The chaotic pattern found at around 4\,Myr grows and doubles in width by age 20\,Myr, then persists 
for the rest of the main-sequence life of the star; it covers more than a factor of 10 
in envelope mass $\Delta M/M_*$\footnote{$\Delta M$ is the envelope mass, measured from the surface}.  
The Mn-Fe-Ni convection zone eventually extends from $\log T=5.10$ to 5.45 in the grey model, which evolves more smoothly thanks to this well mixed, single ``iron-group'' convection zone; that simulation also appears 
as the red curves in Fig\,\ref{melody_star_internals}.
Note that in all three models, homogeneity is also artificially imposed by mixing, in all the layers 
between the surface and the He\,\textsc{ii} convection zone.


In every case, one finds a massive extended region of marginal convective stability, 
that the code tries to monitor in detail, when some average solution might be more appropriate 
physically and easier to use numerically; this complexity prevented the simulations with weak turbulence 
from continuing.  
The low mass loss used did not affect significantly the ``iron-group'' convection zone; higher mass loss 
rates would have to be investigated, but that would still not help solve the atmosphere problem, 
and in the present case, would reduce both P and Fe in the same way, 
unless inhomogeneous mass loss were introduced; this was not attempted since implementing 
a more realistic atmosphere solution must be done first.

\subsubsubsection{Predicted surface composition}

\begin{figure}
\epsscale{0.60}
\plotone{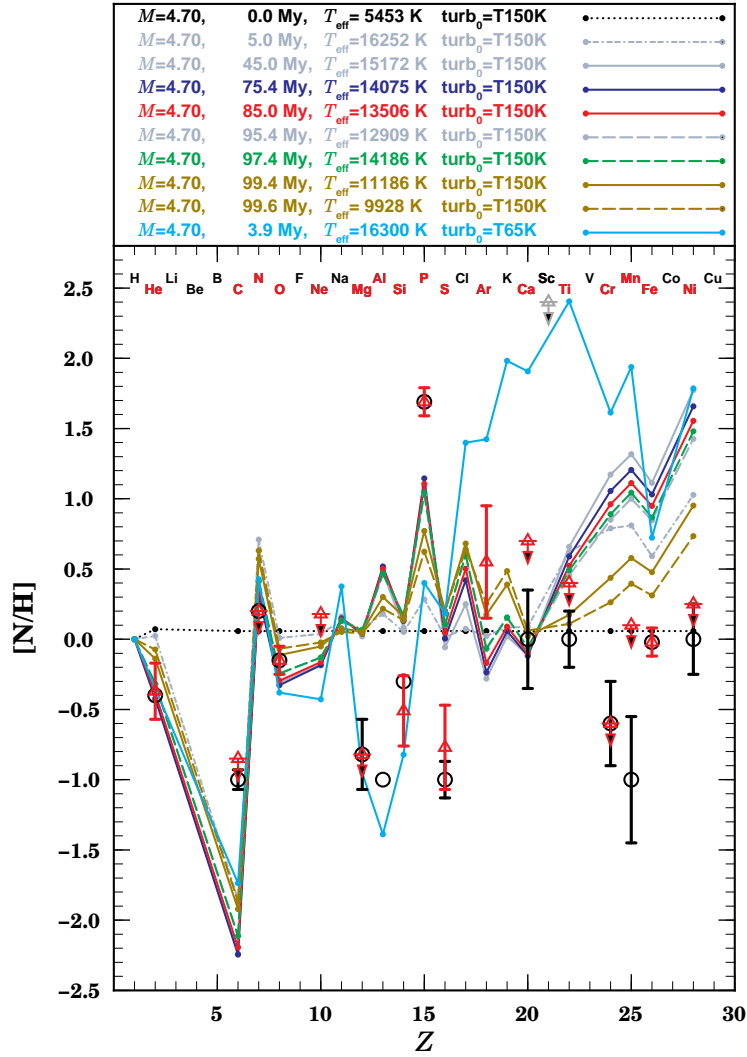}
\caption{Evolution of computed surface abundance patterns in XEVOL models \texttt{T150KD1K-3} (first nine profiles) 
and \texttt{T65KD10-4} (last profile), for star HD\,318101, with $M_*=4.7\protect\Msol$.   
The top legend indicates the age, effective temperature and level of turbulence of each model.
Observations from R.~Monier (\emph{black circles}, private communication) and 
K{\i}l{\i}\c{c}o\u{g}lu et\ al (\emph{red triangles}, present paper) are shown for comparison.
Scandium, measured but not included in the simulations, is shown as a \emph{grey triangle}.
Lines for ages${}>99$\,Myr correspond to the end of the main-sequence and the rapid wiping-out 
of anomalies.  Elements for which there are both measurements and modeled values appear in red at the
top of the plot.  Abundances are relative to the Sun.
\label{M6_vs_xevol1}}
\end{figure}

In Fig.\,\ref{M6_vs_xevol1} 
we compare the predicted surface abundances of all the main elements handled by XEVOL 
for two models of mass $M_*=4.7$\,M$_\odot$, with independent measurements from Monier 
(2014, private communication) and K{\i}l{\i}\c{c}o\u{g}lu et al. (this paper) for the P-rich star HD\,318101.  
Nine different ages are shown from the evolution of model \texttt{T150KD1K-3}
of Fig.\,\ref{melody_star_internals} and \ref{4.7Msol_zc};  these cover the whole main-sequence lifetime of 
the star;  a single age is shown for model \texttt{T65KD10-4} (blue points in Fig.\,\ref{4.7Msol_zc}).

The results do not depend sensitively 
on the stellar mass; a range of masses (we looked at 4.5--5.0\Msol), effective temperatures, 
and ages are possible around these values, but all share very similar abundance profiles, 
with strong positive correlation between P and Fe (see also Fig.\,\ref{melody_star_internals}).  
This can be tracked down to the similarity of the $g_\mathrm{rad}(\mathrm{P})$ and 
$g_\mathrm{rad}(\mathrm{Fe})$ profiles (see Fig.\,\ref{M6_gradM4.7T150K}),  which remain almost identical in shape 
and strength in the deep envelope, throughout the evolution; the only noticable 
difference is in the smoothness of these profiles: the P profile becomes noisier as 
we approach the surface; this is caused by the much smaller number of lines in 
its spectrum.   
The P abundance rises very rapidly then remains constant for almost the whole evolution; the
iron-peak elements (Mn, Fe, Ni) also rise very quickly,  then decrease slowly and steadily until about 90\,Myr.
Unfortunately, this decrease is not quick enough to return Fe to near solar level while P is still high.
As mentioned at the beginning of \S\,\ref{modelingHD318101}, we expect the huge quantities of P and Fe pushed towards the surface to be further stratified in the atmosphere, in different ways, leading to
the observed difference, but this cannot be proved with XEVOL alone.

All elements $\le$\,Ca,  except phosphorus, appear to fit the predicted patterns very well in the age range 45--95\,Myr, 
given the quoted error bars.   As discussed at the end of \S\,{modelingHD318101}, the high P abundance may also agree with models which do not enforce mixing down to the He convection zone (still do be demonstrated).
Some measurements suggest  P might actually be lower in HD\,318101 and closer to the 1.1 mark, in which case the match would be perfect.


\subsubsubsection{Uncertainties in $g_{\mathrm{rad}}(\mathrm{P})$ calculations}

Besides error bar estimates on observations, theoretical models themselves have uncertainties of their own 
in their radiative force calculations; the latter are necessarily propagated to the predicted abundances 
of the models.   Uncertainties about the physics (plasma equation of state, 
core-electron approximations, quantum cross-sections, etc) have been discussed by the authors of the OPAL 
tables.  Uncertainties originating from the use of frequency sampling are \emph{not} known, and may be the 
most important ones, as far as $g_{\mathrm{rad}}$ calculations are concerned.  They are of two types:  1) errors due to insufficient spectrum resolution; 2) errors due to unknown relative positions of lines of different elements.

\citet{leblancmiri2000} studied in a quantitative way the importance of frequency grid resolution on the calculation of $g_{\mathrm{rad}}(A)$, and showed its effects for ten elements $A$ in the OPAL database (but unfortunately not  phosphorus); their Fig.\,4 shows the effect of using low resolution spectra (10\,000 points, non-uniform, optimized for the black body distribution -- see their Appendix A), compared to a uniform $10^6$-point $u$ reference grid covering the same range\footnote{The $10^6$-point calculation was shown, through more exact methods of evaluation (analytical line profile integrations), to be accurate to better than 0.1\,\% on the resulting $g_{\mathrm{rad}}(A)$, for all $A$ studied}; the low resolution and the evaluation temperature (100\,000\,K) are both similar to the conditions in the corresponding XEVOL evaluation of $g_{\mathrm{rad}}(\mathrm{P})$ in the stellar regions where it is maximum.  They verified that phosphorus has very few absorption lines at that temperature in the star, so frequency sampling at the OPAL resolution has a high probability of missing the cores of most of these lines, a situation which reminds one of the case of Lithium.  
We expect  $g_{\mathrm{rad}}(\mathrm{P})$ to carry a similar uncertainty as $g_{\mathrm{rad}}(\mathrm{Li})$, in these stellar layers.    These are errors of the \emph{first type} mentioned above.

Errors of the \emph{second type} occur when the absorption spectrum contains a huge number of lines of some abundant 
element and one needs to compute the light momentum absorbed by another element with very few absorption lines.  
The light available to that second element is blocked by the huge number of lines of the first element absorbing most of the 
momentum; since any of these lines carries a large positional uncertainty (typically $\sim$1\,\%  in the OPAL database), we 
don't know for sure what flux should be available in any of the few lines we are considering.
This kind of uncertainty is not reduced by increasing resolution; it could be reduced only by increasing wavelength accuracy for all elements.

\citet{richermi05} studied the uncertainty on $g_{\mathrm{rad}}(\mathrm{Li})$ resulting from the uncertainty in the relative 
positions of Li and Fe absorption lines in the OPAL database\footnote{This problem is very similar to the one discussed in 
Sections 3 and 4 of \citet{leblancmiri2000}.}.    Using a statistical simulation of possible OPAL-like line placements errors, 
these authors were able to show that $g_{\mathrm{rad}}(\mathrm{Li})$ values at particular temperatures can vary by large 
factors (cf. Fig.\,3 and 4 of \citealt{richermi05}); in the worst cases, these factors could \emph{typically} be
\begin{equation}
\ g_{\mathrm{rad}}(\mathrm{Li})_\mathrm{[true]}= (0.2~\textrm{to}~3.)\times g_{\mathrm{rad}}(\mathrm{Li})_\mathrm{[frequency\ sampling]}\,.\label{sampling_error}
\end{equation}
The resulting diffusion velocity of the element would be uncertain by the same factor; the abundance itself depends exponentially on this velocity in many transients situations, and its uncertainty could be much greater. 
One should repeat this analysis in the case of phosphorus, to clarify the predictions of XEVOL+OPAL calculations for that 
element.   If the same level of uncertainty applies to P in the envelope of HD\,318101, the theoretical predictions of the 
present paper, augmented by this sampling uncertainty factor, will be well within the range of measured values.

\citet{leblancmiri2000} were able to show that for elements such as Fe, 
the huge number of lines available allows the frequency sampling method to produce a 
statistically unbiased estimate of $g_{\mathrm{rad}}(\mathrm{Fe})$.    


\subsubsubsection{Other models}

Differentiated mass loss (inhomogeneous wind) could play a significant role in the evolution of surface abundances of some elements in HD\,318101.   It would be possible to use the present XEVOL model to test this inhomogeneous wind model as a mean of getting rid of the Fe atmospheric excess, while retaining most of the phosphorus at the surface.  Only preliminary work has been done in that direction so far.


\section{Conclusions} \label{conclusions}

The spectra of 44 late-B, A, and F dwarfs of M6 have been synthesized in a uniform manner to derive LTE abundances of up to 20 chemical elements. This is the first extensive abundance study of intermediate mass stars in M6 with stars ranging from 1.4 $M_{\odot}$ up to 4.3 $M_{\odot}$. As we did for other open clusters, we find that  the abundances of several chemical elements exhibit real star-to-star variations in the B and A stars, larger than for the  F stars. The largest spreads occur for [Mg/H], [Mn/H], [Y/H] and [Ba/H] while the smallest are for [Si/H], [Ca/H], [Sc/H], [Ti/H], [Cr/H], [Fe/H]  and [Ni/H].  The derived abundances do not show clear systematic trends with effective temperature, nor with apparent rotational velocities as expected since the timescale of diffusion are much shorter than those of rotational mixing \citep{charbonneaumi91}. The relative abundances [Ti/Fe], [Cr/Fe], [Mn/Fe],  [Ni/Fe], [Y/Fe], and [Ba/Fe] are correlated with that of iron and the relative abundances [C/Fe], [O/Fe], [Si/Fe], and [Sc/Fe] are anticorrelated with [Fe/H] as expected from models including diffusion.
The normal F stars have nearly solar abundances for almost all elements with little scatter around the mean abundances. In these stars, 
Mg, Cr and Y are only marginally overabundant and Ba is overabundant. 
These results agree quite well with the findings of the previous extensive abundance surveys of  \citet{gebetal08}, \citet{gebmon08}, and \citet{gebetal10} for Coma Berenices, the Pleiades and the Hyades.

In the course of this analysis, we have discovered five new chemically peculiar stars of different types in M6. Stars 5, 47, and 69 appear to be a mild-Am, an Am and a Fm star respectively, star 99 is a new HgMn star, and star 20 (HD 318101) is a new He-weak and P-rich star. 

The detailed modeling of the P-rich star HD\,318101 including radiative forces
and different amounts of turbulent diffusion reproduces the overall shape of the abundance pattern 
for this star and the abundances of the light elements He, C and O but not those of 
heavier elements, in particular the iron-peak elements.
Models with the least turbulence reproduce the abundances of the lightest elements 
($\mathrm{Z}<12$) and those with most turbulence reproduce coarsely abundances of elements 
with $\mathrm{Z}>15$ (assuming an age in the 50--90\,Myr range).

The discrepancies between derived and predicted abundances appear to come from the stratification 
of some elements in the atmosphere of that star, a phenomenon which is not handled by the XEVOL code, and requires more realistic (non local) radiative transfer calculations.
However, the manner in which radiative forces change over time in our models for HD\,318101 suggests another solution: a very young model ($\sim\,$50\,Myr) may actually fit most observations (including the phosphorus and Fe-group abundances) if no mixing is introduced far from the convection zones and in particular, at the surface.  Such a model still needs to be investigated with XEVOL, but would certainly be numerically challenging.


\acknowledgments
\textbf{Acknowledgment.} This work and the first author was supported by T\"{U}B\.{I}TAK (The Scientific and Technological Research  Council  of  Turkey) under Program No. 2214 (2011-1).

\appendix

\section{Appendix material}

\subsection{Radial velocities}

The radial velocity of each star was derived by cross-correlating each observed spectrum with an appropriate synthetic spectrum computed for the  \teff{}, \logg{}, \vsini{}, \vmic{}, and chemical composition of the star. The radial velocity obtained for each star is listed in Table \ref{radial}.

We have derived an average cluster radial velocity by weight averaging the radial velocity of the single stars, obtaining $V_r=-8.48\pm0.17$ \kms{}. This value is in agreement with the cluster radial velocity of $-8.27\pm0.45$ \kms{}, given by \citet{frimaj08} derived using six fiducial cluster members.

M6 was observed with the GIRAFFE H09B setting on two different nights, separated by six days, allowing us to look for radial velocity variations which could be due to orbital motions.  Most stars have stable radial velocities, reproducible to within $\pm$  2.74\,\kms{}, which we can consider the intrinsic precision, $\sigma_{v_{rad}}$ of the measurements. We therefore considered stars displaying differences in  $V_r$ much larger than this threshold (ie. larger than 3 $\sigma_{v_{rad}}$) as exhibiting significant radial velocity variations.

The star no.\,99 presents the largest radial velocity variation ($\sim38\,\kms{}$) and is therefore likely to be a binary. We have also detected radial velocity shifts above the uncertainty limit of {2.74\,\kms{}} for stars nos.\,5, 31, 47, 51, 64, and 130. These stars might be single lined spectroscopic binaries (SB1) as the spectral lines of their companion stars are not visible in the observed spectra.

The spectra of the star no.\,33 reveal that this is a double-lined spectroscopic binary (SB2) with components of similar spectral type, because of the similar intensity of the lines of the two stars. The effective temperature derived from Geneva photometry indicated an early-F spectral type. We determined the projected rotational velocities of the two stars using unblended  well separated iron lines, obtaining 30$\pm$10\,\kms{} and 35$\pm$10\,\kms{}. The maximum $V_\mathrm{r}$ difference between the two stars was 170$\pm$10 \kms{}.

\subsection{Non-LTE effects and notes on individual elements} \label{nonlte}

In this section, we discuss possible departures from LTE which could affect the abundances of helium, carbon, oxygen, magnesium, silicon, calcium and scandium.. 

\textbf{Helium:} The helium abundances derived for late-B type stars can be regarded as solar within the uncertainties, except for the star no.\,20, which is classified as a He-weak P-rich star. The He abundances were derived from the synthesis of the He I 4713\AA{}, 4921\AA{}, and 5026\,\AA{} lines which have very accurate \loggf{} values. 
\citet{przetal11} have shown that for stars cooler than  $\teff{} \simeq 22\,000$\,K 
two of these lines, He I 4713 and 5016\,\AA{}, are the least affected by non-LTE effects. 
We also checked \citet{nieprz07} where these transitions are studied in LTE and NLTE. The recent analysis of helium lines in the HgMn star $\kappa$ Cancri by \citet{maznie14} reveals a significant strengthening of a few He I lines prone to isotopic shifts, eg. $\lambda$ 4921 \AA\ and $\lambda$ 6678 \AA\ which we have not used to derive the helium abundance in the new HgMn star we found in this study.

The B-type stars analysed here are much cooler than 22\,000\,K. We believe that the non-LTE corrections for He are within about the abundance uncertainties (i.e., $\sim0.2$ dex) and we did not correct our LTE determinations.  Note that for the hottest CP4 star no.\,20, the correction is negative, and this would make helium even more deficient.

\textbf{Carbon and Oxygen:} We derived the C abundance of the A and F-type stars mainly from four C I lines having quality B and C  log$gf$ values. All stars have nearly solar C abundance, but for the two Am stars which are deficient in carbon. We checked the departures from LTE for the considered C I lines using the analysis of the spectrum of Vega (A0V) in \citet{przetal01a}. The largest departure (i.e., $-0.04$\,dex) occurs for the C I 4772\,{\AA{}} line. Given the fairly large uncertainties on the C abundance, the departure from LTE for carbon can be considered to be negligible.

We used several O I lines with \loggf{} values of quality C+, in particular the multiplet at 5329\,{\AA{}}, to derive the oxygen abundance. Oxygen is either normal or only slightly under-abundant with respect to the solar abundance \citep{gresau98} for the B and A-type stars, while it is underabundant for the star no. 47 (Am). As for carbon, we used \citet{przetal00} to check the departures LTE: non-LTE effects for the lines we analysed do not exceed $-0.03$\,dex and can therefore be neglected..

\textbf{Magnesium and Silicon:} There are four relatively strong Mg I lines in the observed spectra of late-A and F stars: Mg I 4702.991, 5167.321, 5172.684, and 5183.604\,{\AA{}}. The first three of these lines have  oscillator strengths of quality B+, while the last one is of quality  A.  We noticed 
that the Mg I 4702.991 line systematically yields abundances which are less than those derived from the other three lines.
 We suppose that the low abundances derived from this line may be due to an erroneous oscillator strength. We thus preferred to synthesize the triplet Mg I lines at  5170 \AA{} only to derive the magnesium abundance. \citet{przetal01b} showed that the Non-LTE correction is $-0.06$\,dex for Mg I 5172.684, while it is $-0.13$\,dex for Mg I 5183.604, for Vega (A0V). This suggests that our derived Mg abundances may be  slightly overestimated. This non-LTE correction will decrease the slightly large Mg abundances of the stars, and bring them to solar-like values. For the B and early-A stars, several Mg II lines are available for synthesis, such as Mg II 4739.593, 4739.709, 5264.220, and 5264.364. \citep{przetal01b} showed that the non-LTE effects on Mg II lines (between 4500-5800 \AA{}) can be considered as negligible in the case of of Vega. We note that the Mg abundance derived from these lines for B and early-A type members are indeed found to be solar \citep{gresau98}.

The Si abundance was derived from several Si II lines in the spectra, which are of quality B or D, and was found solar for most members with $\teff{}<12\,000$\,K. The abundances, however, seem to be reduced for the hottest stars. \citet{wed01}, who performed a non-LTE abundance analysis for Si found that the Non-LTE correction of Si is $+0.05$\,dex for Vega for the lines we analysed. Przybilla et al. (2011) have shown that lines of Si II are not influenced by non-LTE effects in stars cooler than about 15000 K. \citet{bailan13} performed a few checks on the Si II lines of late B-type stars which support this result.

\textbf{Calcium:} Calcium and scandium are crucial elements since they should be underabundant in Am stars. We could only use a limited number (about four) of lines, for both elements to derive their abundances, as many other lines were too weak to appear in the spectra. 

Each member was found to have nearly solar Ca abundances, except for nos.\,5 and 71 which are slightly under-abundant, and for no.\,69 which is slightly overabundant ($+0.25$\,dex). No.\,5 displaying the lowest [Ca/H] is a mild Am star. The unexpected deficiency of Ca for No. 71, which is a chemically normal star, is probably due to the difficulties to model Ca lines, since this star has \vsini{} of 135\,\kms{}. The Ca enrichment in No.\,69 is unusual as this star is a most-likely a CP1 (Fm) star. In order to find out whether our results were affected by LTE approximation, we used \citet{sitetal14}'s results, including the temperature dependent non-LTE corrections of the Ca II 5001\,{\AA{}} line. According to them, non-LTE corrections of this line is negative and does not exceed $-0.03$\,dex for $5000\Kelvin < \teff{} < 8700\Kelvin$ (also for \logg{}=4 and [Fe/H]=0.0). The corrections, however, become positive for higher temperatures, and departures considerably increase for $\teff{} > 9000 \Kelvin$, and reach about to +0.19\,dex at \teff{}=10\,000\Kelvin. We thus conclude that the derived Ca abundances for members earlier than A3 type might be slightly underestimated in our LTE approximation.


\subsubsection{Comments on particular stars}
 
We discuss here i) the stars which stand out from the Main Sequence in the HR diagram of M6, ii) the newly found Chemically Peculiar stars.
 
The stars no.\,25, 118 and HD\,318103 were found to be more luminous,
than the other cluster members of similar effective temperatures in the HR diagram of M6 while stars nos.\,41 and 95 were found to be less luminous . Unknown binarity can also be the reason of these offsets. The derived radial velocity of the star no.\,25 ($-6.76\pm1.68$\kms{}) differs much from what was derived by \citet[$-24.73\pm4.63$\kms{}]{frimaj08}. This difference may arise from a faint companion star whose lines are invisible in the spectra of the star. Time series observations are clearly needed for no.\,25 for confirmation of its duplicity, and for no.\,118 and HD\,318103 to find out whether they are binaries or not.

We present below the abundance pattern of the newly detected chemically peculiar stars.

\subsubsection{NGC\,6405\,5 (HD\,318091)} The abundance pattern of NGC\,6405\,5 indicates that this star is a mild-Am star. Both Ca and Sc are underabundant while Cr and Ba are overabundant relative to the Sun. The star presents also the lowest C abundance among the analyzed stars. The abundances of Mg, Si, Ti, Fe, Ni, Y are nearly solar within their error limits. Its radial velocity variation also suggests that this star is most likely a binary.  

\subsubsection{NGC\,6405\,20 (HD\,318101)} \label{sec:NGC_6405_20} This star is the hottest and the brightest star among our program stars. 
The membership of this star was confirmed by \citet{frimaj08} from its three-dimensional motion.
The abundance pattern of NGC\,6405\,20 indicates that this star is most likely a He-weak\,P[Ga?] (CP4) star. Its effective temperature and low rotational velocity support this classification (see Kurtz 2000). Although we detected a marked overabundance of P and Xe, we failed to identify Ga lines in this spectral region. A spectrum covering the region 3000-4000\,\AA{} would be very valuable to check the existence of Ga lines. 

\subsubsection{NGC\,6405\,47 (CD-32\,13109)} This is the star with the lowest \vsini{}. This star presents clear Am chemical peculiarities: overabundances of Cr, Mn, Fe, Ni, Y, and Ba and underabundances of C, O and Sc. The Ca abundance is solar within the uncertainties. The radial velocity measurements also indicate that the star is most likely in a binary system. 

\subsubsection{NGC\,6405\,69 (GSC\,07380-01211)} The abundance pattern of NGC\,6405\,69 shows that the star is a Fm star. It is slightly enriched in Ca, Cr, Fe, Y and Ba. These peculiarities are not as large as those of classical CP1 stars and there is no deficiency of Sc. We did not find any evidence of duplicity for this star from its radial velocity variations.

\subsubsection{NGC\,6405\,99 (HD\,318126)} The abundance pattern of NGC\,6405\,99 indicates that this star is most likely a HgMn star. We detected overabundances of Ti, Cr, Mn, Ni, Zr, Y, Ba and Hg. We could only synthesize Hg I 5769.593 line to derive the Hg abundance . A spectrum including the Hg II 3983.93 \AA\ lines would be very valuable to determine more accurately its Hg abundance. Its binarity detected from radial velocity variations also supports its HgMn classification.

%
%
%
%

\bibliography{tk}

\clearpage




\begin{deluxetable}{lllll} 
\tabletypesize{\scriptsize}
\tablecolumns{4}
\tablewidth{0pc}
\tablecaption{Age, distance, color excess and metallicity of M6}
\tablehead{
\colhead{Reference} & \colhead{Age}   & \colhead{Distance}    & \colhead{Color Excess} & \colhead{Metallicity} \\
\colhead{} & \colhead{(Myr/log\textit{t}[s])}   & \colhead{(pc)}    & \colhead{$E(B-V)$} & \colhead{[Fe/H]} }
\startdata

\citet{rohetal59} & 100 / 8.00 & $630\pm50$ & 0.156 & \nodata \\ 
\citet{bec61} & \nodata & 505 & 0.1 & \nodata \\ 
\citet{egg61}$^\mathrm{a}$ & \nodata & 501 & 0.13 & \nodata \\
\citet{johetal61} & \nodata & $630\pm90^\mathrm{b}$ & 0.16 & \nodata \\ 
\citet{tal65} & \nodata & $480\pm40$ & \nodata & \nodata \\ 
\citet{lin68} & 50 / 7.70 & \nodata & \nodata & \nodata \\ 
\citet{becfen71} & \nodata & 445 & 0.17 & \nodata \\ 
\citet{vle74} & 100 / 8.00 & $450\pm50$ & 0.15 & \nodata \\
\enddata
\tablenotetext{a}{reported via \citet{ant72}}
\tablenotetext{b}{error adopted from the uncertainties of their given distance modulus}
\tablenotetext{c}{coverted value from $E(B_2-V_1)=0.135$ (see Section \ref{tefflogg})}
\tablenotetext{d}{value calculated from their given their given distance modulus (8.3 mag)}

\label{intro}
\end{deluxetable}

\begin{deluxetable}{cccccccccc} 
\tablecolumns{4}
\tablewidth{0pc}
\tablecaption{Observation log}
\tablehead{
\colhead{Date} & \colhead{Setting}   & \colhead{MJD}    & \colhead{Exposure}  \\
\colhead{} & \colhead{}   & \colhead{(start of exp. time)}    & \colhead{time (s)} }
\startdata
30.05.2007 & H09B & 54250.34380149 & 410 \\ 
           &      & 54250.34941879 & 410 \\ 
           &      & 54250.35476998 & 410 \\ 
05.06.2007 & H09B & 54256.12558588 & 410 \\ 
           &      & 54256.13093846 & 410 \\ 
           &      & 54256.13629534 & 410 \\ 
           & H11  & 54256.14208523 & 290 \\
           &      & 54256.14605750 & 290 \\ 
           &      & 54256.15002295 & 290 \\ 
           & LR3  & 54256.15521383 & 120 \\
           &      & 54256.15721080 & 120 \\ 
\enddata
\label{log}
\end{deluxetable}

\begin{deluxetable}{lccccccccc} 
\tabletypesize{\scriptsize}
\tablecolumns{9}
\tablewidth{0pc}
\tablecaption{List of selected targets}
\tablehead{
\colhead{}    &  \colhead{Rohlfs et}    &  
\colhead{} & \multicolumn{2}{c}{Membership probability} & 
\multicolumn{2}{c}{Radial velocity (\vrad)} &  \colhead{HR} &  \colhead{} \\
\colhead{Star} & \colhead{al. (1959)}   & \colhead{\textit{V}}    & \colhead{D2006} &
\colhead{FM2008}    & \colhead{FM2008}   & \colhead{This study}    & \colhead{Diagram} &\colhead{S/N}  \\
\colhead{} & \colhead{id. no.}   & \colhead{}    & \colhead{(per cent)} &
\colhead{(per cent)}    & \colhead{(\kms{})}   & \colhead{(\kms{})}    & \colhead{criterion} &
\colhead{}}
\startdata
\object[HD 160093]{HD\,160093} & 33 & 11.29 & \nodata & 97.1 & 39.23$\pm$2.58 & (SB2) & Yes & 221/156/123 \\ 
\object[HD 160167]{HD\,160167} & 115 & 8.56 & 61 & 1.1 & -1.53$\pm$5.94 & -8.21$\pm$3.07 & Yes & 482/447/396 \\ 
\object[HD 160188]{HD\,160188} & \nodata & 9.31 & \nodata & \nodata & \nodata & -4.34$\pm$2.80 & Yes & 319/400/355 \\
\object[HD 160222]{HD\,160222} & 97 & 8.50 & 52 & \nodata & \nodata & 6.76$\pm$18.18 & Yes & 478/616/516 \\
\object[HD 160259]{HD\,160259} & 28 & 8.73 & 4 & 4.4 & -71.71$\pm$5.09 & 0.02$\pm$6.17 & Yes & 427/510/401 \\ 
\object[HD 160260]{HD\,160260} & 70 & 8.32 & 61 & \nodata & \nodata & -14.38$\pm$9.44 & Yes & 573/686/539 \\
\object[HD 160297]{HD\,160297} & 41 & 8.84 & 39 & \nodata & \nodata & -5.11$\pm$2.59 & No & 479/512/403 \\ 
\object[HD 160298]{HD\,160298} & 94 & 8.60 & 3 & 10.4 & -65.24$\pm$5.68 & -10.51$\pm$1.70 & Yes & 522/564/444 \\
\enddata
\label{fundamental}
\end{deluxetable}


\begin{deluxetable}{lccccccc}
\tabletypesize{\scriptsize}
\tablecolumns{6}
\tablewidth{0pc}
\tablecaption{Log $L/L_\odot$, log \teff{}, $M/M_{\odot}$ and fractional age ($\tau$) for the members}
\tablehead{
\colhead{Star} & \colhead{Id no.}   & \colhead{log $L/L_\odot$}    & \colhead{log \teff{}} & \colhead{$M/M_{\odot}$}    & \colhead{$\tau$} }
\startdata
HD\,160167 & 115 & 2.14 & 4.086 & $3.12\pm0.22$ & $0.29\pm0.05$ \\ 
HD\,160188 & \multicolumn{1}{l}{\nodata} & 1.72 & 4.029 & $2.53\pm0.19$ & $0.17\pm0.03$ \\ 
HD\,160222 & 97 & 2.27 & 4.134 & $3.64\pm0.31$ & $0.43\pm0.09$ \\ 
HD\,160259 & 28 & 2.07 & 4.086 & $3.13\pm0.31$ & $0.29\pm0.08$ \\ 
HD\,160260 & 70 & 2.37 & 4.146 & $3.83\pm0.38$ & $0.48\pm0.12$ \\ 
HD\,160297 & 41 & 1.14 & 4.049 & $2.57\pm0.13$ & $0.17\pm0.02$ \\ 
HD\,160298 & 94 & 2.17 & 4.107 & $3.33\pm0.23$ & $0.34\pm0.06$ \\ 
HD\,160392 & 17 & 1.91 & 4.061 & $2.86\pm0.30$ & $0.23\pm0.06$ \\
\enddata
\label{fractional}
\end{deluxetable}

\begin{deluxetable}{lcclrlccr} 
\tabletypesize{\scriptsize}
\tablecolumns{10}
\tablewidth{0pc}
\tablecaption{Fundamental parameters for the programme stars}
\tablehead{
\colhead{}    &  \colhead{Rohlfs et}    & \multicolumn{2}{c}{Photometric}  & 
\multicolumn{2}{c}{Final} &  \colhead{} &  \colhead{} \\
\colhead{Star} & \colhead{al. (1959)}   & \colhead{\teff{}}    & \colhead{\logg{}} &
\colhead{\teff{}}    & \colhead{\logg{}}   & \colhead{\vmic{}}    &
\colhead{\vsini{}}   & \colhead{Sp. T.} \\
\colhead{} & \colhead{id. no}   & \colhead{(K)}    & \colhead{(dex)} &
\colhead{(K)}    & \colhead{(dex)}   & \multicolumn{2}{c}{(\kms{})}  & \colhead{}}
\startdata
HD\,160093 & 33 & 6892 & 4.35 & $7400\pm500$ & \nodata & \multicolumn{2}{c}{(spectroscopic binary)} & \nodata \\
HD\,160167 & 115 & 13032 & 4.19 & $12200\pm300$ & $4.00\pm0.10$ & 0 & $120\pm10$ & B8 \\ 
HD\,160188 & \nodata & (12632) & \nodata & $10700\pm300$ & $4.10\pm0.10$ & 0 & $300\pm20$ & B9 \\ 
HD\,160222 & 97 & 13433 & 4.03 & $13600\pm400$ & $4.00\pm0.15$ & 0 & $310\pm20$ & B7 \\ 
HD\,160259 & 28 & 11793 & 4.35 & $12200\pm500$ & $4.15\pm0.10$ & 0 & $240\pm20$ & B8 \\ 
HD\,160260 & 70 & 13967 & 3.99 & $14000\pm500$ & $4.10\pm0.10$ & 0 & $250\pm25$ & B6 \\ 
HD\,160297 & 41 & 11200 & 4.20 & $11200\pm200$ & $4.20\pm0.10$ & 0 & $280\pm20$ & B9 \\ 
HD\,160298 & 94 & 12961 & 4.04 & $12800\pm300$ & $4.00\pm0.10$ & 0 & $290\pm20$ & B7 \\
\enddata
\tablenotetext{a}{The star IDs are listed in columns 1 and 2, the \teff{} and \logg{} derived from Geneva 7-color photometry in column 3, the final adopted \teff{} and \logg{} in col. 4, the derived microturbulent velocity, radial velocity and rotational velocity and a spectral type estimated from \teff{} in columns 5, 6, 7 and 8}
\label{physical}
\end{deluxetable}

\begin{deluxetable}{lccrl}
\tablecolumns{11}
\tablewidth{0pc}
\tablecaption{Atomic data and their references}
\tablehead{
Elem.  & $\lambda$ (\AA{}) & \textit{E}$_\mathrm{low}$(cm$^{-1}$) &$\log gf$ &Ref.  \\ }
\startdata
\multicolumn{5}{l}{Helium}  \\
\ion{He}{1} & 4713.139 & 169086.859 & $-1.277$ & 1(N) \\ 
\ion{He}{1} & 4713.156 & 169086.938 & $-1.498$ & 1(N) \\ 
\ion{He}{1} & 4713.376 & 169087.922 & $-1.975$ & 1(N) \\ 
\ion{He}{1} & 4921.931 & 171135.000 & $-0.443$ & 1(N) \\ 
\ion{He}{1} & 5015.678 & 166277.547 & $-0.820$ & 1(N) \\ 
\multicolumn{5}{l}{Carbon} \\ 
\ion{C}{1} & 4771.742 & 60393.140 & $-1.866$ & 2(N) \\
\enddata

\tablerefs{K: via Kurucz's gfhyperall.dat, V: via VALD2, N: via NIST Atomic Spectra Database, VK: atomic data from VALD2 and \loggf{} fractions for $hfs$ from Kurucz's gfhyperall.dat, 1: \citet{wiefuh09}, 2: \citet{wieetal96}, 3: \citet{kelpod08a}, 4: \citet{kurpey75}, 5: \citet{kelpod08b}, 6: \citet{miletal71}, 7: \citet{wieetal69}, 8: \citet{hib88}, 9: \citet{podetal09}, 10: \citet{the89}, 11: \citet{lawdak89}, 12: \citet{picetal02}, 13: \citet{maretal88}, 14: \citet{fuhwie06} and/or \citet{brietal91}, 15: \citet{raauyl98}, 16: \citet{wiclaw97}, 17: \citet{fuhwie05}, 18: \citet{cowcor83}, 19: \citet{ljuetal06}, 20: \citet{ryasmi89}, 21: \citet{wiemar80}, 22: \citet{cur04}}
\label{atomic}
\end{deluxetable}


\begin{deluxetable}{lccccccccccc}
\tabletypesize{\tiny}
\tablecolumns{6}
\tablewidth{0pc}
\tablecaption{Log(X/H) absolute abundances for the M6 dwarfs }
\tablehead{Star & Star no. & He	& C & O & Mg & Si & P & S & Ar & Ca & Sc}
\startdata
HD 160093 & 33 & \nodata & \nodata & \nodata & \nodata & \nodata & \nodata & \nodata & \nodata & \nodata & \nodata \\ 
HD 160167 & 115 & -1.19 & \nodata & \nodata & \nodata & -4.68 & \nodata & \nodata & \nodata & \nodata & \nodata \\ 
HD 160188 & \nodata & -1.38 & \nodata & \nodata & \nodata & -4.75 & \nodata & \nodata & \nodata & \nodata & \nodata \\ 
HD 160222 & 97 & -1.03 & \nodata & \nodata & \nodata & -4.41 & \nodata & \nodata & \nodata & \nodata & \nodata \\ 
HD 160259 & 28 & -1.31 & \nodata & \nodata & \nodata & -4.47 & \nodata & \nodata & \nodata & \nodata & \nodata \\ 
HD 160260 & 70 & -0.94 & \nodata & \nodata & \nodata & -4.27 & \nodata & \nodata & \nodata & \nodata & \nodata \\ 
HD 160297 & 41 & \nodata & \nodata & \nodata & \nodata & -4.37 & \nodata & \nodata & \nodata & \nodata & \nodata \\ 
HD 160298 & 94 & -1.20 & \nodata & \nodata & \nodata & -4.34 & \nodata & \nodata & \nodata & \nodata & \nodata \\
\enddata
\label{abuntab}
\end{deluxetable}


\begin{deluxetable}{lccccccc}
\tabletypesize{\tiny}
\tablecolumns{6}
\tablewidth{0pc}
\tablecaption{Abundances for the M6 Dwarfs with respect to the cosmic abundance standards \citep{nieprz12}}
\tablehead{Star & Star no. & [He/H] & [C/H] & [O/H] & [Mg/H] & [Si/H] & [Fe/H]}
\startdata
HD 160093 & 33 & \nodata & \nodata & \nodata & \nodata & \nodata & \nodata \\ 
HD 160167 & 115 & -0.18 & \nodata & \nodata & \nodata & -0.18 & -0.26 \\ 
HD 160188 & \nodata & -0.37 & \nodata & \nodata & \nodata & -0.25 & 0.09 \\ 
HD 160222 & 97 & -0.02 & \nodata & \nodata & \nodata & 0.09 & \nodata \\ 
HD 160259 & 28 & -0.30 & \nodata & \nodata & \nodata & 0.03 & 0.12 \\ 
HD 160260 & 70 & 0.07 & \nodata & \nodata & \nodata & 0.23 & -0.02 \\ 
HD 160297 & 41 & \nodata & \nodata & \nodata & \nodata & 0.13 & 0.06 \\ 
HD 160298 & 94 & -0.19 & \nodata & \nodata & \nodata & 0.16 & 0.11 \\
\enddata
\label{cosmictab}
\end{deluxetable}    
    

\begin{deluxetable}{ccccccc}
\tablecolumns{8}
\tablewidth{0pc}
\tablecaption{Uncertainties of the abundances for late B stars}
\tablehead{
\colhead{Species} & \colhead{$\sigma_{\teff{}}$}   & \colhead{$\sigma_{\logg{}}$}    & \colhead{$\sigma_{\vmic}$} & \colhead{$\sigma_{\mathrm{norm.}}$}    & \colhead{$\sigma_{\mathrm{tot}}$} & \colhead{$\sigma_{\mathrm{max.}}$} \\
 & \colhead{$\pm300$\,K}   & \colhead{$\pm0.10$\,dex}    & \colhead{$0/1$\,\kms{}} & \colhead{$\pm$(1/SNR)}    & \colhead{}   & \colhead{} }
\startdata
\ion{He}{0} & 0.10 & 0.12 & 0.00 & 0.01 & 0.16 & \nodata  \\
\ion{O}{0} & 0.02 & 0.02 & 0.04 & 0.02 & 0.05 & 0.25  \\
\ion{Mg}{0} & 0.10 & 0.02 & 0.00 & 0.01 & 0.11 & 0.16  \\
\ion{Si}{0} & 0.02 & 0.01 & 0.03 & 0.03 & 0.05 & 0.10 \\
\ion{P}{0} & 0.01 & 0.02 & 0.03 & 0.01 & 0.04 & \nodata  \\
\ion{S}{0} & 0.04 & 0.04 & 0.01 & 0.04 & 0.07 & \nodata  \\
\ion{Ar}{0} & 0.05 & 0.04 & 0.03 & 0.19 & 0.20 & \nodata  \\
\ion{Ti}{0} & 0.15 & 0.08 & 0.07 & 0.05 & 0.19 & 0.21  \\
\ion{Cr}{0} & 0.15 & 0.11 & 0.08 & 0.06 & 0.21 & 0.58 \\
\ion{Mn}{0} & 0.05 & 0.04 & 0.04 & 0.05 & 0.09 & \nodata \\
\ion{Fe}{0} & 0.00 & 0.01 & 0.06 & 0.07 & 0.09 & 0.24 \\
\ion{Ni}{0} & 0.09 & 0.01 & 0.00 & 0.16 & 0.19 & \nodata \\
\ion{Y}{0} & 0.29 & 0.19 & 0.08 & 0.06 & 0.36 & \nodata \\
\ion{Zr}{0} & 0.16 & 0.01 & 0.02 & 0.03 & 0.16  & \nodata \\
\ion{Xe}{0} & 0.08 & 0.10 & 0.00 & 0.08 & 0.15 & \nodata \\
\ion{Ba}{0} & 0.11 & 0.03 & 0.14 & 0.07 & 0.20 & \nodata \\
\ion{Hg}{0} & 0.06 & 0.03 & 0.09 & 0.48 & 0.50 & \nodata \\
\enddata
\label{Bstarerr}
\end{deluxetable}

\begin{deluxetable}{ccccccc}
\tablecolumns{8}
\tablewidth{0pc}
\tablecaption{Uncertainties of the abundances for A stars}
\tablehead{
\colhead{Species} & \colhead{$\sigma_{\teff{}}$}   & \colhead{$\sigma_{\logg{}}$}    & \colhead{$\sigma_{\vmic}$} & \colhead{$\sigma_{\mathrm{norm.}}$}    & \colhead{$\sigma_{\mathrm{tot}}$}   & \colhead{$\sigma_{\mathrm{max.}}$}  \\
 & \colhead{$\pm150$\,K}   & \colhead{$\pm0.10$\,dex}    & \colhead{$\pm0.20$\,\kms{}} & \colhead{$\pm$(1/SNR)}    & \colhead{}   }
\startdata
\ion{C}{0} & 0.10 & 0.03 & 0.01 & 0.08 & 0.13 & 0.14  \\
\ion{O}{0} & 0.01 & 0.01 & 0.00 & 0.04 & 0.04 & 0.07  \\
\ion{Mg}{0}  & 0.09 & 0.03 & 0.01 & 0.02 & 0.10 & 0.16  \\
\ion{Si}{0} & 0.01 & 0.01 & 0.01 & 0.02 & 0.03 & 0.06  \\
\ion{Ca}{0} & 0.06 & 0.02 & 0.01 & 0.06 & 0.09 & 0.09  \\
\ion{Sc}{0} & 0.06 & 0.01 & 0.02 & 0.06 & 0.08 & 0.10  \\
\ion{Ti}{0} & 0.07 & 0.01 & 0.02 & 0.01 & 0.07 & 0.11  \\
\ion{Cr}{0} & 0.08 & 0.00 & 0.01 & 0.01 & 0.08 & 0.12  \\
\ion{Mn}{0} & 0.04 & 0.00 & 0.01 & 0.04 & 0.05 & 0.19  \\
\ion{Fe}{0} & 0.07 & 0.01 & 0.02 & 0.01 & 0.08 & 0.13  \\
\ion{Ni}{0} & 0.12 & 0.03 & 0.00 & 0.03 & 0.13 & 0.18  \\
\ion{Y}{0} & 0.12 & 0.00 & 0.00 & 0.02 & 0.12 & 0.23  \\
\ion{Ba}{0} & 0.16 & 0.07 & 0.08 & 0.04 & 0.19 & 0.19  \\
\enddata
\label{Astarerr}
\end{deluxetable}

\begin{deluxetable}{ccccccccc}
\tablecolumns{8}
\tablewidth{0pc}
\tablecaption{Uncertainties of the abundances for F stars}
\tablehead{
\colhead{Species} & \colhead{$\sigma_{\teff{}}$}   & \colhead{$\sigma_{\logg{}}$}    & \colhead{$\sigma_{\vmic}$} & \colhead{$\sigma_{\mathrm{norm.}}$}    & \colhead{$\sigma_{\mathrm{tot}}$}  & \colhead{$\sigma_{\mathrm{max.}}$} & \\
 & \colhead{$\pm150$\,K}   & \colhead{$\pm0.30$\,dex}    & \colhead{$\pm0.20$\,\kms{}} & \colhead{$\pm$(1/SNR)}    & \colhead{} & \colhead{} }
\startdata
\ion{C}{0} & 0.03 & 0.09 & 0.00 & 0.04 & 0.10 & 0.25  \\
\ion{Mg}{0} & 0.14 & 0.16 & 0.01 & 0.02 & 0.21  & 0.23   \\
\ion{Si}{0} & 0.04 & 0.07 & 0.01 & 0.16 & 0.18 & 0.19  \\
\ion{Ca}{0} & 0.04 & 0.02 & 0.05 & 0.02 & 0.07 & 0.12  \\
\ion{Sc}{0} & 0.04 & 0.11 & 0.02 & 0.08 & 0.13 & 0.28  \\
\ion{Ti}{0} & 0.06 & 0.07 & 0.02 & 0.04 & 0.10 & 0.13  \\
\ion{Cr}{0} & 0.05 & 0.05 & 0.02 & 0.05 & 0.09 & 0.14 \\
\ion{Mn}{0} & 0.09 & 0.01 & 0.03 & 0.04 & 0.11 & 0.15  \\
\ion{Fe}{0} & 0.06 & 0.01 & 0.03 & 0.05 & 0.08 & 0.13  \\
\ion{Ni}{0} & 0.03 & 0.01 & 0.01 & 0.05 & 0.06 & 0.12  \\
\ion{Zn}{0} & 0.07 & 0.04 & 0.02 & 0.09 & 0.13 & 0.22  \\
\ion{Y}{0} & 0.07 & 0.08 & 0.05 & 0.02 & 0.12 & 0.28  \\
\ion{Ba}{0} & 0.10 & 0.05 & 0.09 & 0.03 & 0.15 & 0.22  \\
\enddata
\label{Fstarerr}
\end{deluxetable}


\begin{deluxetable}{ccccccccc}
\tablecolumns{9}
\tablewidth{0pc}
\tablecaption{Mean abundances with respect to the Sun \citep{gresau98} and star-to-star abundance variations}
\tablehead{
\colhead{}    & \multicolumn{2}{c}{A-type stars} & \multicolumn{2}{c}{A-type stars}    &  \multicolumn{2}{c}{F-type stars} & \multicolumn{2}{c}{F-type stars} \\
\colhead{} & \multicolumn{2}{c}{}   & \multicolumn{2}{c}{(except CP stars)}    & \multicolumn{2}{c}{} & \multicolumn{2}{c}{(except CP stars)} \\
\colhead{Element} & \colhead{$\mathrm{[\langle{}X/H\rangle{}]}$}   & \colhead{$\sigma$([X/H])}    & \colhead{$\mathrm{[\langle{}X/H\rangle{}]}$} & $\sigma$([X/H]) & \colhead{$\mathrm{[\langle{}X/H\rangle{}]}$} & $\sigma$([X/H]) & \colhead{$\mathrm{[\langle{}X/H\rangle{}]}$} & $\sigma$([X/H])}
\startdata

C & $-0.10$ & 0.13 & $-0.05$ & 0.06 & 0.02 & 0.05 & 0.02 & 0.06 \\
O & $-0.15$ & 0.10 & $-0.12$ & 0.11 & \nodata & \nodata & \nodata & \nodata \\
Mg & 0.18 & 0.17 & 0.22 & 0.18 & 0.13 & 0.07 & 0.12 & 0.08 \\
Si & 0.09 & 0.11 & 0.10 & 0.13 & 0.08 & 0.04 & 0.09 & 0.03 \\
Ca & $-0.12$ & 0.21 & $-0.08$ & 0.12 & 0.07 & 0.09 & 0.04 & 0.07 \\
Sc & 0.02 & 0.28 & 0.11 & 0.08 & 0.02 & 0.07 & 0.01 & 0.06 \\
Ti & $-0.03$ & 0.12 & $-0.07$ & 0.08 & 0.03 & 0.07 & 0.01 & 0.06 \\
Cr & 0.25 & 0.21 & 0.12 & 0.11 & 0.16 & 0.09 & 0.13 & 0.05 \\
Mn & 0.24 & 0.18 & 0.15 & 0.17 & 0.13 & 0.14 & 0.08 & 0.10 \\
Fe & 0.04 & 0.15 & $-0.04$ & 0.08 & 0.10 & 0.07 & 0.07 & 0.03 \\
Ni & 0.12 & 0.23 & $-0.07$ & 0.03 & 0.04 & 0.06 & 0.02 & 0.05 \\
Y & 0.33 & 0.26 & 0.24 & 0.19 & 0.19 & 0.16 & 0.12 & 0.08 \\
Ba & 0.90 & 0.51 & 0.46 & 0.31 & 0.57 & 0.10 & 0.53 & 0.06 \\

\enddata
\label{startostar}
\end{deluxetable}


\begin{deluxetable}{lrrc} 
\tablecolumns{4}
\tablewidth{0pc}
\tablecaption{Coefficients for the abundances vs. [Fe/H]}
\tablehead{
\multicolumn{4}{c}{$Y = a[Fe/H] + b$} \\
\colhead{$Y$}    & \colhead{$a$} & \colhead{$b$} &  \colhead{$R^2(^*)$}}
\startdata
[C/Fe] & $-1.20$ & $-0.04$ & $0.704$ \\ 
{[O/Fe]} & $-1.60$ & $-0.10$ & $0.998$ \\ 
{[Mg/Fe]} & $-1.04$ & $0.11$ & $0.369$ \\ 
{[Si/Fe]} & $-1.19$ & $0.11$ & $0.927$ \\ 
{[Ca/H]} & $0.65$ &  $-0.07$ & $0.297$ \\ 
{[Sc/Fe]} & $-1.76$ & $0.04$ & $0.722$  \\ 
{[Ti/H]} & $0.66$ & $-0.05$ & $0.717$  \\ 
{[Cr/H]} & $1.01$ & $0.11$ & $0.702$ \\ 
{[Mn/H]} & $1.02$ & $0.06$ & $0.719$ \\ 
{[Ni/H]} & $0.66$ & $-0.05$ & $0.835$ \\ 
{[Y/H]} & $1.72$ & $0.04$ & $0.768$ \\ 
{[Ba/H]} & $2.50$ & $0.41$ & $0.585$ \\  
\enddata
\tablenotetext{*}{$R^2:$ Coefficient of determination}
\label{coeff}
\end{deluxetable}


\begin{deluxetable}{lcccccccc} 
\tabletypesize{\scriptsize}
\tablecolumns{6}
\tablewidth{0pc}
\tablecaption{Radial velocity variations}
\tablehead{
\colhead{}    & \colhead{Rohlfs et} & \colhead{MJD}    &  \colhead{MJD} & \colhead{} & \colhead{Radially}\\
\colhead{Star} & \colhead{al. (1959)}   & \colhead{54250.3494}    & \colhead{54256.1309} &
\colhead{Difference}    & \colhead{variable?} \\
\colhead{} & \colhead{id. no}   & \colhead{\vrad\ (\kms{})}    & \colhead{\vrad\ (\kms{})} & \colhead{$\mid\Delta\vrad\mid$\ (\kms{})}}
\startdata
HD\,160167 & 115 & $-10.39$ & $-11.49$ & $1.10$ & No  \\ 
HD\,160392 & 17 & $-8.09$ & $-9.43$ & $1.35$ & No \\ 
HD\,318091 & 5 & $-7.29$ & $-3.07$ & $4.22$ & Yes \\ 
HD\,318093 & 25 & $-5.29$ & $-7.30$ & $2.02$ & No \\ 
HD\,318094 & 42 & $-4.69$ & $-7.29$ & $2.60$ & No \\
HD\,318099 & 29 & $-6.09$ & $-8.08$ & $1.99$ & No \\ 
HD\,318101 & 20 & $-8.39$ & $-7.98$ & $0.41$ & No \\ 
HD\,318103 & \nodata & $-8.79$ & $-6.47$ & $2.32$ & No \\
\enddata
\label{radial}
\end{deluxetable}

\end{document}